%% file: NN_eEDM_PRL_v2.tex
\documentclass[aps,prd,preprintnumbers,superscriptaddress,nofootinbib,floatfix,twocolumn,notitlepage]{revtex4-2}
\usepackage{graphicx}  
\usepackage{dcolumn}   
\usepackage[table]{xcolor} 
\usepackage{makecell}   
\usepackage{bm}        
\usepackage{epsfig,amsmath,amssymb,verbatim,mathrsfs,array,layout,textcomp,amssymb,latexsym,slashed}
\usepackage{xcolor}
\usepackage[colorlinks=true,citecolor=blue,urlcolor=blue,linktocpage=true,
linkcolor=blue]{hyperref}
\usepackage[utf8]{inputenc}
\usepackage{multirow}
\usepackage{xspace}
\usepackage{booktabs} 
\usepackage{siunitx} 
\usepackage{tabularx} 
\usepackage{physics} 
\usepackage{dsfont} 
\usepackage{physunits} 

\newcommand{\cL}{\mathcal{L}}

\newcommand{\cO}{\mathcal{O}}

\newcommand{\cE}{\mathcal{E}}

\newcommand{\eg}{\textit{e.g.}}
\newcommand{\ie}{\textit{i.e.}}

\newcommand{\alphaNP}{\alpha_{\phi}}
\newcommand{\mNP}{m_{\phi}}
\newcommand{\gS}{g_{\rm S}}
\newcommand{\gP}{g_{\rm P}}

\newcommand{\eEDM}{\textit{e}EDM}

\newcommand{\req}{r_{\rm eq}}

\newcommand{\GeV}{\textrm{GeV}}
\newcommand{\TeV}{\textrm{TeV}}

\newcommand{\be}{\begin{equation}}
\newcommand{\ee}{\end{equation}}
\newcommand{\bea}{\begin{eqnarray}}
\newcommand{\eea}{\end{eqnarray}}

\newcommand{\HfF}{$\text{HfF}^+\xspace$}
\newcommand{\ThF}{$\text{ThF}^+\xspace$}

\definecolor{light_blue}{rgb}{0.15, 0.35, 0.9}

\newcommand{\mysec}[1]{\textbf{#1}}

\definecolor{Ecolor}{HTML}{008000}
\definecolor{Bcolor}{HTML}{c8ab37}
\definecolor{edm}{HTML}{c83771}

\definecolor{onlyP}{HTML}{003f5c}
\definecolor{onlyN}{HTML}{bc5090}
\definecolor{bothPN}{HTML}{ffa600}

\newcommand{\caseA}{\textcolor{onlyP}{(a)}}
\newcommand{\caseB}{\textcolor{onlyN}{(b)}}
\newcommand{\caseC}{\textcolor{bothPN}{(c)}}

\usepackage[inkscapeformat=png]{svg}

\begin{document}

\title{Constraining CP Violating Nucleon-Nucleon Long Range Interactions in Diatomic eEDM Searches}
 
\author{Chaja Baruch}
\email{chajabaruch@campus.technion.ac.il}
\affiliation{Physics Department, Technion -- Israel Institute of Technology, Haifa 3200003, Israel}

\author{P.~Bryan Changala}
\email{bryan.changala@cfa.harvard.edu}
\affiliation{Center for Astrophysics $\vert$ Harvard \& Smithsonian, Cambridge, MA 02138, USA}

\author{Yuval Shagam}
\email{yush@technion.ac.il}
\affiliation{Schulich Faculty of Chemistry, The Helen Diller Quantum Center and the Solid State Institute,
Technion -- Israel Institute of Technology, Haifa, 3200003, Israel}

\author{Yotam Soreq}
\email{soreqy@physics.technion.ac.il}
\affiliation{Physics Department, Technion -- Israel Institute of Technology, Haifa 3200003, Israel}

\begin{abstract}
The searches for CP violating effects in diatomic molecules, such as \HfF and ThO, are typically interpreted as a probe of the electron's electric dipole moment~(\eEDM{}), a new electron-nucleon interaction, and a new electron-electron interaction. 
However, in the case of a nonvanishing nuclear spin, a new CP violating nucleon-nucleon long range force  will also affect the measurement, providing a new interpretation of the \eEDM{} experimental results.  
Here, we use the \HfF \eEDM{} search and derive a new bound on this hypothetical interaction, which is the most stringent from terrestrial experiments in the 1\eV-10\eV[k] mass range.   
These multiple new physics sources motivate independent searches in different molecular species for CP violation at low energy that result in model independent bounds, which are insensitive to cancellation among them. 
\end{abstract}

\maketitle

\mysec{Introduction.}
Notwithstanding its great success, the Standard Model~(SM) is not a complete description of Nature and should be extended by physics beyond the Standard Model~(BSM), which is well motivated both by observational evidence and strong theoretical arguments, see \eg~\cite{EuropeanStrategyforParticlePhysicsPreparatoryGroup:2019qin}.
New physics~(NP) sources of CP violation~(CPV) naturally appear in a variety of extensions of the SM and may be related to Baryogenesis.  
Low energy BSM searches, \eg~\cite{Safronova:2017xyt}, can probe these effects. 
In particular, CPV searches are sensitive to multiple NP effects, \eg{} electric dipole moments~(EDMs)~\cite{Pospelov:1997uv,Ginges:2003qt,Pospelov:2005pr,Engel:2013lsa, Chupp:2017rkp, Gaul:2023hdd}.
Focusing on electron EDM~(\eEDM) searches in diatomic molecules, NP CPV can arise not only in the form of the \eEDM{}, but also as a new CPV electron-nucleon ($eN$) or electron-electron ($ee$) interaction~\cite{Stadnik:2017hpa,Fleig:2018bsf,Prosnyak:2023duq,Flambaum:2019ejc}.
To date, the most stringent \eEDM{} bound is $|d_e| <4.1 \times 10^{-30}e\,\cm$~\cite{Roussy:2022cmp,Caldwell:2022xwj}, assuming no other CPV sources. 
This can be translated to new physics at the scale of $\cO(10\,\TeV)$. 

In this Letter, we point out that a new CPV nucleon-nucleon~($NN$) long-range force mediated by a spin-0 particle contributes to the \eEDM{} frequency channel and that this effect is probed by measurements in diatomic molecules, in the case where one or both of the nuclei of the diatomic molecule have nonzero spin.
This presents a set of models that are probed by reintepreting the current \eEDM{} results.
In addition to the \eEDM{}, CPV can arise from long range forces between electrons and nuclei, such that there are four NP CPV sources, namely $d_e$, $eN$, $ee$ and $NN$. 
To constrain these four CPV sources, at least four independent measurements are required. 

The three most sensitive \eEDM{} searches in molecules are the JILA \HfF{} search~\cite{Roussy:2022cmp}, the ACME ThO search~\cite{ACME:2018yjb}, and the Imperial College London YbF search~\cite{Hudson:2011zz,Kara:2012ay}.
Only the first and last include nuclei with nonvanishing spins and are sensitive to $NN$. 
We utilize these three searches to derive novel bounds on three NP CPV sources, \ie~$d_e$, $eN$ and $NN$.
We note that $ee$ contributes to $d_e$ and directly through the measured frequency channel. 
The interaction can be also probed by atomic EDM searches~\cite{Stadnik:2017hpa} with decent precision. 
We neglect it here for simplicity and leave this comparison for future work. 
Our result is the most constraining bound on $NN$ from terrestrial experiments, improving current constraints by up to 6 orders of magnitude in the $1\,\eV-10\,\keV$ mass range. 
Because the YbF \eEDM{} bound is weaker by $\cO(100)$ compared to the other \eEDM{} searches sensitive to $NN$, the upcoming \ThF experiment~\cite{Ng:2022fzm}, which also contains one nucleus with a nonzero spin, is further motivated and will lead to a 100-fold improved sensitivity.  
Astrophysical bounds from stellar cooling~\cite{Bottaro:2023gep,Hardy:2016kme} and neutron stars~\cite{Buschmann:2021juv} are stronger by 2-3
orders of magnitude, see also~\cite{OHare:2020wah}. 
However, these astrophysical bounds are subject to large systematic uncertainties and moreover, can be avoided in certain models, see below. 

\begin{figure*}[t]
    \centering
    \includegraphics[width=\textwidth]{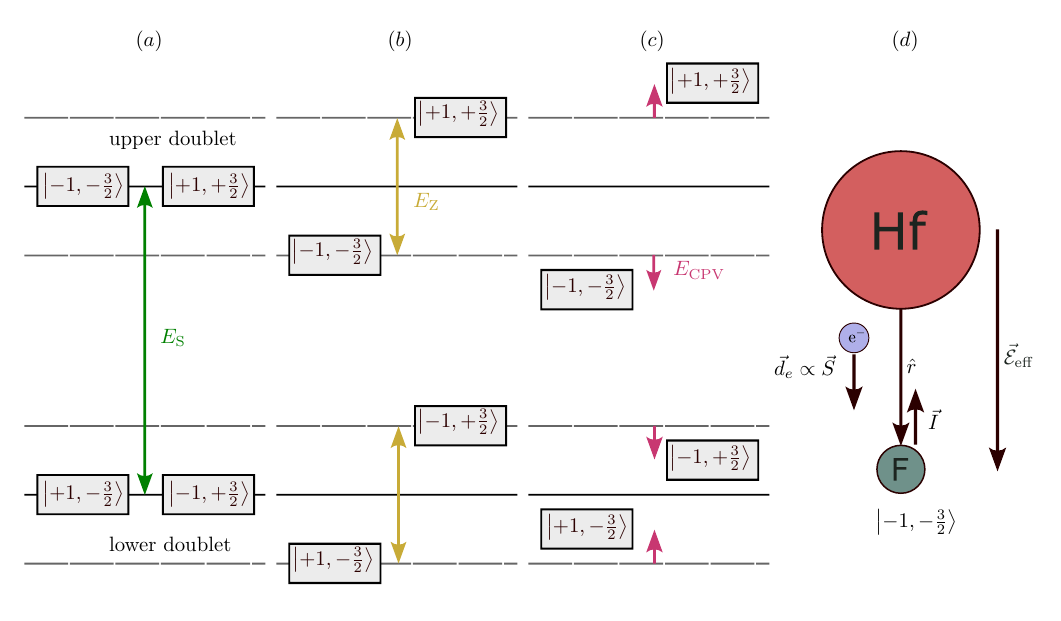}
    \caption{The splitting of the energy levels, $\ket{\Omega,m_F}$, is shown when applying an electric field in~(a), a parallel electric and magnetic field in~(b) and the additional shift due to  $E_{\rm CPV}$, \ie~\eEDM{}, $NN$ and $eN$, is shown in~(c). 
    The relative directions are illustrated in~(d).
    See main text for details.}
    \label{fig:lvl_splitting}
\end{figure*}

\mysec{Long range CPV force.}
We consider a new spin-0 particle, $\phi$, with mass $\mNP$ and both scalar and pseudo-scalar couplings to fermions. 
The effective couplings between $\phi$, the nucleons, $N=n,p$, and the electrons are given by 
\begin{align}
    \label{eq:Lint}
    \cL_{\rm int} 
    \subset   
    \sum_{\psi=e,p,n}\!\!\!\phi \left( \gS^\psi  \overline{\psi} \psi  
    + i \gP^\psi \overline{\psi}\gamma_5 \psi  \right) \, ,
\end{align}
and can be mapped to UV models. 
For example, see~\cite{Moody:1984ba} and~\cite{DiLuzio:2023lmd,DiLuzio:2023edk} for a recent review on the CPV axion.
If $\phi$ is the QCD axion the expected CPV is too small to be observed, see \eg~\cite{Okawa:2021fto,Dekens:2022gha}.
Another example is relaxion models~\cite{Graham:2015cka,Gupta:2015uea,Flacke:2016szy}, which induce a CPV light scalar as a result of mixing with the SM-Higgs boson. 

The effective couplings from Eq.~\eqref{eq:Lint} are constrained both by terrestrial experiments and astrophysical observations, see \eg~\cite{OHare:2020wah,AxionLimits}. 
For $m_\phi \lesssim 10\,\keV$, the most stringent bounds on the effective coupling come from stellar cooling $\gS^N < 6.5\times 10^{-13}$~\cite{Bottaro:2023gep}, see also~\cite{Hardy:2016kme}, from cooling of hot neutron stars $(\gP^n,\gP^p) < (1.3 ,1.5) \times 10^{-9}$~\cite{Buschmann:2021juv} and from SN1987 $\gP^N<6.0\times 10^{-10}$~\cite{Carenza:2019pxu}.  
It has been suggested that the SN1987 bound faces substantial uncertainties, casting doubt on its robustness~\cite{Bar:2019ifz}.
All of these astrophysical bounds can be avoided in models that are subject to environmental effects, see \eg~\cite{Burrage:2016bwy,Masso:2005ym,Jaeckel:2006xm,DeRocco:2020xdt,Budnik:2020nwz,Bloch:2020uzh} and discussion in~\cite{OHare:2020wah}.  
Additionally, the constraints we set in this work are weaker only by 2-3 orders of magnitude compared to the astrophysical constraints.
Moreover, at the one-loop level, the scalar and pseudoscalar proton couplings contribute to the scalar-photon and pseudoscalar-photon couplings. 
Following Ref.~\cite{Escribano:2020wua} to translate the strongest bounds on these photon couplings,  we obtain a bound of $\cO(10^{-16})$ on $g_S^p g_P^p$. 
In this case, globular cluster bounds~\cite{Dolan:2022kul} are subject to the same ambiguity as mentioned above. 

Considering terrestrial experiments, the most stringent bounds arise from the proton and neutron EDMs, $|d_p|<2.1\times 10^{-25}e\,\cm$~\cite{Sahoo:2016zvr} and $|d_n|<1.8\times 10^{-26}e\,\cm$~\cite{Abel:2020pzs,Pendlebury:2015lrz}, respectively.
Following Ref.~\cite{DiLuzio:2023edk} and assuming $m_\phi\ll\GeV$, the bounds on the effective couplings of Eq.~\eqref{eq:Lint} are  $\gS^p \gP^p < 8.4\times 10^{-10}$ and $\gS^n \gP^n < 1.0\times 10^{-10}$. 
Additional bounds via pion nucleon coupling lead to $\gS^{n,p} \gP^{n,p} < 10^{-9}-10^{-11}$~\cite{Mantry:2014zsa}.
For $m_\phi\lesssim \eV$ there are very stringent bounds from searches with macroscopic objects \eg~\cite{Lee:2018vaq,Jenke:2012ju,Tullney:2013wqa,Guigue:2015fyt,Feng:2022tsu}, see discussion in~\cite{OHare:2020wah}. 
The scalar and pseudoscalar couplings can be separately probed by CP conserving observables, such as molecular vibrational modes \eg~\cite{SchillerHD1,SchillerHD2} and rare Kaon decays \eg~\cite{Goudzovski:2022vbt}.  
The  combined strongest laboratory bounds on the scalar coupling, from neutron scattering~\cite{Nesvizhevsky:2007by,Kamiya:2015eva,Haddock:2017wav,Heacock:2021btd}, and pseudoscalar coupling, from molecular HD~\cite{Ledbetter:2012xd}, together give a constraint of $\gS^n \gP^n < 1.0\times 10^{-16}$ for $m_\phi \lesssim 1 \eV$ and $\gS^n \gP^n < 1.0\times 10^{-9} - 10^{-10}$ for the keV range. 
For additional bounds, see \eg~\cite{Mostepanenko:2020lqe}.
Bounds on the CPV scalar-photon coupling can be found in \eg~\cite{Gorghetto:2021luj}. They can be reinterpreted as a bound on the proton coupling and the terrestrial bounds are found to be weaker than other relevant bounds in the \eV[k] range. 

The effective monopole-dipole potential between two nuclei $i$ and $j$ is given by~\cite{Dobrescu:2006au,JacksonKimball:2014vsz,Fadeev:2018rfl}
\begin{align}
    \label{eq:Vps}
    V_{\rm SP}(r)
=    \alphaNP^{ij} \frac{\vec{\sigma}_j\cdot\hat{r}}{2\overline{m}_N} 
    \left(\frac{1}{r} + m_\phi \right) \frac{e^{-r m_\phi}}{r}\,,
\end{align}
where $\vec{\sigma}_j$ are the Pauli matrices that follow the spin of the valance nucleon, $\overline{m}_N = 939\,\eV[M] $ is the average nucleon mass and $\vec{r}$ is the internuclear axis of the molecule. 
The NP interaction strength is defined 
\begin{align}
    \label{eq:alphaNP}
    \alphaNP^{ij} \equiv 
    -\frac{1}{4\pi} 
    (Z^i \gS^p + N^i \gS^n)
    (B^j_p \gP^p  + B^j_n \gP^n) \,,
\end{align}
where $Z^i\,(N^i)$ is the number of protons\,(neutrons) in nucleus $i$.
The relation between the nucleus and the nucleon spins, as well as the proton-neutron mass difference, is encoded in $B^j_{n,p}$~\cite{JacksonKimball:2014vsz, Flambaum:2006ip,Stadnik:2014xja}, see Supplemental Material. 
Since for molecules $r\sim 10^{-10}\,\m$, a $\phi$ with $m_\phi \lesssim 2\,\keV$ will induce a long range force at the molecular scale.

\mysec{Nucleon-nucleon long range CPV force in diatomic molecules.}
\begin{table*}[t]
   \def\arraystretch{1.3}
    \begin{center}
    \begin{tabular}{l|c|c|c|c|c}
    \hline \hline
    Molecule &    
    $\req$ [\AA] & 
    $W^{ij}_{NN}\,[\Hz[\mu]]$ & 
    $W^{ij}_{d_e}\,[\Hz[\mu]/e\,\m[c]]$ & $W^{ij}_{eN}\,[\Hz[\mu]]$ & 
    $E_{\rm CPV}^{ij}\,[\Hz[\mu]]$   \\        
\hline                                                        
  $^{180}$\HfF    &   1.81~\cite{PhysRevA.76.030501}  & $1.55 \times 10^{17}$ & $5.55 \times 10^{30} $~\cite{Fleig:2018bsf} & $1.16\times10^{22}$~\cite{Stadnik:2017hpa}  & $-7.3 \pm 11.9$~\cite{Roussy:2022cmp} \\
 $^{232}$ThO &   1.84~\cite{Khristenko1998}  & -  & $1.91 \times 10^{31}  $~\cite{Fleig:2018bsf} & $1.03\times10^{22}$~\cite{Stadnik:2017hpa}   & $81.2 \pm 77.2$~\cite{ACME:2018yjb}    \\
 $^{174}$YbF     &  2.02~\cite{Khristenko1998}  &  $1.24 \times 10^{17}$   & $-3.12\times 10^{30}$~\cite{Fleig:2018bsf} & $2.21\times10^{22}$~\cite{Stadnik:2017hpa}   & $748.7 \pm 1810.5$~\cite{Hudson:2011zz}   \\  
\hline
  $^{232}$\ThF     &  1.99 \cite{Barker2012} & $1.28 \times 10^{17}$ & $9.02 \times 10^{30}$~\cite{PhysRevA.91.042504} &  & $0\pm 5 $ (projection)~\cite{Ng:2022fzm}   \\
  \hline
  \hline
   \end{tabular}  
 \caption{Summary of \eEDM{} measurements and the parameters relevant to computing the NN coupling effect. 
 $W^{ij}_{eN}$ and $W^{ij}_{NN}$ are given in the limit of $\mNP\ll\keV$.  
 } 
   \label{tbl:edm exps}
   \end{center}
 \end{table*}
Here we briefly describe the recently reported \eEDM{} measurement using trapped \HfF{} molecules~\cite{Roussy:2022cmp,Caldwell:2022xwj}, which we use as an example to illustrate the effect.  
The measurement is focused on the $^3\Delta_1$, $J=1$, $F=3/2$ science state. 
Leveraging the $^3\Delta_1$ state's $\Omega-$doubling, the state is polarized by a rotating electric field to form states of well-defined orientation, called the upper and lower doublets, denoted by $\Omega m_F=\pm 3/2$, see Fig.~\ref{fig:lvl_splitting}\,(a). 
Here $\Omega=\pm1$ denotes the quantum number of the projection of the electronic angular momentum, $J=L+S=1$, onto the internuclear axis in the molecule frame and $m_F$ is the quantum number of the projection of the total angular momentum including nuclear spin, $F=J+I$, in the rotating-frame. 
A bias magnetic field is aligned or antialigned with the electric field to lift the remaining degeneracy between the $m_F=\pm 3/2$ states of both $\Omega m_F=\pm 3/2$ doublets, Fig.~\ref{fig:lvl_splitting}\,(b).  
In total there are four relevant states, $\ket{\Omega,m_F}=\ket{\pm1,\pm3/2}$.

The energy of each state is a sum of a common energy shift ($E_0$), a state dependent Stark shift ($E_{\rm S}$), Zeeman shift ($E_{\rm Z}$) and a CPV shift ($E_{\rm CPV}$), which includes the \eEDM{} and other NP effects.
In total, we can write
\begin{align}
    \label{eq:E}
    E_{\Omega,m_F}
=   E_0 + E_{\rm S} + E_{\rm Z} + E_{\rm CPV} \, .
\end{align}
The signs of $E_{\rm S}$, $E_{\rm Z}$ and $E_{\rm CPV}$ are proportional to the signs of $\Omega m_F$, $m_F B_0$ and $\Omega m_F^2$, respectively, where $B_0$ is the magnetic field. 
The sign of the Stark shift follows the orientation of the molecule axis relative to the rotating-frame, which is defined by the rotating electric field.
The Zeeman shift follows the orientation of the electron spin relative to the rotating-frame which is aligned with the magnetic field. 
The CPV is proportional to both $\Omega m_F \times m_F$, i.e. to the product of the external electric and magnetic fields. 

In an example iteration of the experiment, the molecules are prepared in the $\ket{\pm1,+3/2}$ states. 
A Ramsey spin-precession measurement is conducted between the $m_F=+3/2$ and $-3/2$ states for the two doublets with the results measured separately. 
Here $E_0$ and $E_{\rm S}$ are common to the $m_F=\pm3/2$ states and are canceled to leading order where residual effects are suppressed.
For a positive $E_{\rm CPV}$, the two states $m_F=\pm3/2$ with opposite $\Omega$ move closer together\,(farther apart) in lower, $\Omega m_F=-\frac{3}{2}$\,(upper, $\Omega m_F=\frac{3}{2}$) doublets, Fig.~\ref{fig:lvl_splitting}\,(c). 
The doubly differential measurement between the results for the two doublets as well as the aligned and anti-aligned magnetic fields is used to isolate the CPV effects. 
Thus, under $(B_0,m_F,\Omega m_F)\to(-B_0,-m_F,-\Omega m_F)$ (or in other words, the $f_{DB}$ channel) we are left with 
\begin{align}
     \label{eq:CPresult}
     E_{\rm CPV}
=&   \frac{(E_{+1,+\frac{3}{2}} \!-\! E_{-1,-\frac{3}{2}}) 
    - (E_{-1,+\frac{3}{2}} \!-\! E_{+1,-\frac{3}{2}})}{4} \, ,  
\end{align}
and the sign of $B_0$ is same as the sign of $m_F$. 
The measurements from other experimental switches are averaged to suppress sources of systematic uncertainty, which are not written in Eq.~\eqref{eq:E} for simplicity.  

The principle of the measurement illustrated above is common to all \eEDM{} experiments such as \HfF{}, ThO, and \ThF where $\Omega-$doubling is in effect. 
In other cases such as YbF, the electric field direction is switched instead of comparing doublet states. 
In all these cases, fully-stretched states of the total angular momentum $F$ are used to orient the \eEDM{} with respect to the molecule, whose strong internal electric fields polarize to the electron. 
This effective electric field, $\vec{\cE}_{\rm eff}$, points along the $\hat{r}$ direction, \ie{} $\hat{r}\sim \Omega \, m_F$.
This means that any nonzero nuclear spin, namely that of $^{19}$F ($I = 1/2$), will be oriented with both the \eEDM{} and the internuclear axis.

Next, we describe why this measurement is also sensitive to the CPV $NN$ interaction.
In the \eEDM{} experiment, the polarizing electric field $E_{\rm rot}$ is parallel or anti-parallel to the magnetic field $B_{\rm rot}$. 
Moreover, for $F=3/2$, the nuclear spin of fluorine must be oriented with $J$. 
Thus, for the fully stretched $m_F$ states the nuclear spin points along the internuclear axis and against the electron spin.

More explicitly, the energy shift of the \eEDM{} $\propto \langle\vec{\cE}_{eff}\cdot \vec{d}_e\rangle\propto \langle\hat{r}\cdot \vec{S}\rangle\propto \langle\hat{r}\cdot \vec{I}\rangle \propto V_{\rm PS}$ of Eq.~\eqref{eq:Vps} for each of the 4 states with $I$ the nuclei spin.
Therefore, the effect of Eq.~\eqref{eq:Vps} behaves as $V_{\rm PS} \propto \langle \hat{r} \cdot \vec{I} \rangle \propto \Omega m_F \times  m_F \propto \Omega m_F^2$. 
This means that for the upper\,(lower) doublet the states $m_F=\pm 3/2$ move closer together\,(farther apart) in energy due the $V_{\rm PS}$ interaction for $\gP^N\gS^N >0$, which is exactly the ($\mathcal{T}$-violating) behavior of the \eEDM{} and their contributions to the $E_{\rm CPV}$ energy cannot be distinguished.
The splitting of the energy levels when applying electromagnetic fields is shown in Figs.~\ref{fig:lvl_splitting}\,(a)-(b).
Therefore, in addition to $d_e$, and CPV in $eN$ and $ee$ interactions~\cite{Stadnik:2017hpa,Fleig:2018bsf,Prosnyak:2023duq}, the diatomic \eEDM{} searches can be affected also from $NN$ CPV long range forces as captured by the potential of Eq.~\eqref{eq:Vps}.

The four relevant CPV sources in the diatomic molecule $ij$ can be written as
\begin{align}
    \label{eq:omegaij}
    E_{\rm CPV}^{ij}
=   W^{ij}_{d_e} d_e 
    + W^{ij}_{eN} \alphaNP^{ie} 
    + W^{ij}_{ee} \alphaNP^{ee}
    + W^{ij}_{NN} \alphaNP^{ij}  \, , 
\end{align}
where the $eN$, $ee$ and $NN$ contributions are functions of $\mNP$, $\alphaNP^{ie}\equiv(Z^i \gS^p + N^i \gS^n)\gP^e/4\pi$ and  $\alphaNP^{ee}\equiv\gS^e\gP^e/4\pi$.
The $W^{ij}_{NN} \alphaNP^{ij}$ is the new CPV nucleon-nucleon interaction due to the potential in  Eq.~\eqref{eq:Vps}.
The relevant $W^{ij}$'s are summarized in Table~\ref{tbl:edm exps} with other experimental values, see also~\cite{Prosnyak:2023duq} for  $W^{\text{\HfF}}_{eN}$ and $W^{\text{\HfF}}_{ee}$.
Since the $ee$ interaction can be directly probed by atomic systems, \eg~\cite{Stadnik:2017hpa}, and contributes also to the \eEDM{}, we neglect it below, but it is straightforward to include. 
We note that the $eN$ interaction can also be probed by atomic systems, see \eg~\cite{Stadnik:2017hpa,Dzuba:2018anu}.

The nucleon-nucleon interaction function can be estimated by using first-order perturbation theory 
\begin{align}
    W^{ij}_{NN}
=   \frac{\left\langle V_{\rm SP} \right\rangle}{\alphaNP^{ij}}
    \approx
    \frac{\langle\vec{\sigma}_j \cdot\hat{r}_{\rm eq} \rangle}{2\overline{m}_N} 
    \left(\frac{1}{\req} + m_\phi \right) 
    \frac{e^{-\req m_\phi}}{\req} \, ,
\end{align}
where $\req$ is the equilibrium distance between the two nuclei and $\langle\vec{\sigma}_j \cdot\hat{r}_{\rm eq} \rangle$ is the nuclear spin expectation value on the internuclear axis for the state. In the fully stretched state (maximal angular momentum), $\langle\vec{\sigma}_j \cdot\hat{r}_{\rm eq} \rangle = 1$.
Although we have taken the internuclear distance to be fixed to its equilibrium value, we have verified, using the Morse anharmonic potential, that the correction from consideration of the nuclear vibrational wave functions of the ground state is at the few percent level for most of the relevant $\mNP$ range and at most $\cO(1)$ for the high masses. 

To further emphasize the importance of the $NN$ effect and encourage future CPV measurements in new molecules, we estimate the possible limit that will likely be set in the mature \ThF measurement, assuming that a value consistent with zero is found.
To predict the limit for \ThF, we use the description in~\cite{Ng:2022fzm} to achieve a conservative estimate of the expected precision assuming a shot noise limited measurement. 
Reference~\cite{Ng:2022fzm} predicts a coherence time of 20\,s and Ref.~\cite{kbthesis} reports 10\,ions counted per shot on the side of the fringe in both doublets in a comparable measurement system. 
We also assume the fringe contrast is 0.55 as was found for \HfF~\cite{Roussy:2022cmp} due to the similarities between the two systems. 
If a total of 600 h of data as in Ref.~\cite{Roussy:2022cmp} are taken using the conveyor belt of ion traps architecture mentioned in Ref.~\cite{Ng:2022fzm,kbthesis}, which would allow for a 10\,Hz repetition rate, a total of $10^8$ ions will be detected neglecting dead times and early time phase measurements. 
Using these parameters we predict a statistical uncertainty of $\sim5\,\mu $Hz which we use for the projection here, see Table~\ref{tbl:edm exps}.

\mysec{Results}
%
\begin{table*}[]
   \def\arraystretch{1.5}
    \begin{center}
    \begin{tabular}{l|c|c|c||c|c}
\hline 
\hline
     &  \multicolumn{3}{c||}{\makecell{Measured \\(\HfF, ThO, YbF)}} & \multicolumn{2}{c}{\makecell{Projection \\ (\HfF, ThO, YbF, \ThF)}} \\
    \hline
     &  $NN$ & $NN,\, d_e$  & $NN,\, d_e,\, eN$  
     &  $NN$ & $NN,\, d_e$  
     \\       
\hline
    \caseA~$g_S^p g_P^p$ & 
    $2.8\times10^{-17}$ & $7.6\times 10^{-17}$ & $3.8\times 10^{-15}$ & 
    $8.8 \times 10^{-18}$ & $7.1\times 10^{-17}$  \\
    \caseB~$g_S^n g_P^n$ & 
    $2.5\times10^{-16}$ & $6.8\times10^{-16}$ & $3.1\times 10^{-14}$ & 
    $7.6 \times 10^{-17}$ & $6.6\times 10^{-16}$  \\
    \caseC~$g_S^{p,n} g_P^{p,n}$ & 
    $1.2\times10^{-17}$ & $3.3\times 10^{-17}$ & $1.6\times10^{-15}$ & 
    $3.7\times 10^{-18}$ & $3.1\times10^{-17}$  \\
\hline                                                        
    $d_e$  & 
    \cellcolor{gray!25} 
    & $1.1 \times 10^{-29}$ & $3.6 \times 10^{-28}$ 
    & \cellcolor{gray!25} 
    & $8.5\times10^{-30}$   \\
\hline
    \caseA~$g_S^p g_P^e$ & 
    \cellcolor{gray!25} & \cellcolor{gray!25} 
    & $8.3\times10^{-20}$ 
    & \cellcolor{gray!25} & \cellcolor{gray!25} 
    \\
    \caseB~$g_S^n g_P^e$ & 
    \cellcolor{gray!25} & \cellcolor{gray!25} 
    & $5.4\times10^{-20}$ 
    & \cellcolor{gray!25} & \cellcolor{gray!25} 
    \\
    \caseC~$g_S^{p,n} g_P^e$ & 
    \cellcolor{gray!25} & \cellcolor{gray!25} 
    & $3.3\times10^{-20}$ 
    & \cellcolor{gray!25} & \cellcolor{gray!25} 
     \\
\hline
\hline
   \end{tabular}  \\
 \caption{Summary of results, with 90\% C.L. The columns indicate which interactions are switched on, see Eq.~\eqref{eq:omegaij}. The rows indicate the relevant couplings, for: only proton coupling~\caseA, only neutron coupling~\caseB, or equal proton and neutron couplings~\caseC. For the \eEDM{} the results for \caseA, \caseB, \caseC{} differ by less than 10 percent.} 
   \label{tbl:results}
   \end{center}
 \end{table*}
 %
\begin{figure}[t]
    \centering
    \includegraphics[width=\columnwidth]{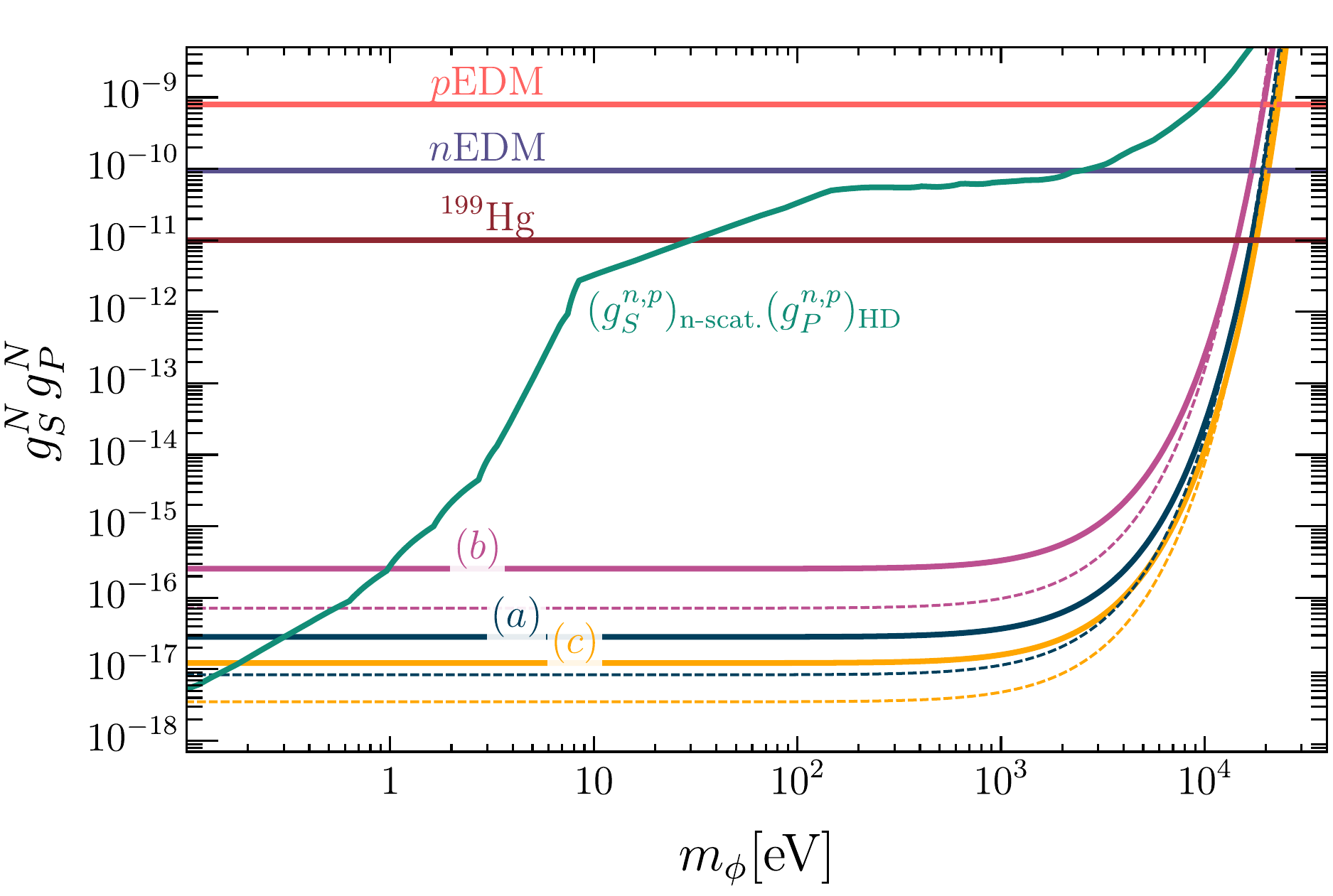}
    \caption{
    The bounds on long-range nucleon-nucleon CPV interaction for: only proton interaction \caseA, only neutron interaction \caseB, and both proton and neutron interaction \caseC.
    Bounds from neutron and proton EDMs~\cite{DiLuzio:2023edk}, from $^{199}$Hg~\cite{Mantry:2014zsa}, as well as the combined strongest laboratory bounds on the scalar coupling, from neutron scattering~\cite{Nesvizhevsky:2007by,Kamiya:2015eva,Haddock:2017wav,Heacock:2021btd}, and pseudoscalar coupling, from molecular HD~\cite{Ledbetter:2012xd}, are also shown. 
    Projected bounds due to a future \ThF measurement are plotted in dashed lines.}
    \label{fig:onlyNN}
\end{figure}
 To set bounds on the couplings under consideration in this work, we use the measured frequency shift of \HfF, ThO and YbF.  
Additionally, we give projected bounds when considering the future \ThF measurement. 
This improves the sensitivity for the scenario in which all interactions are turned on by nearly 2 orders of magnitude. 

We first consider the case of only the $NN$ interaction and set the other CPV sources to zero.
As there is only one unknown, namely $\gS^N \gP^N$, it is sufficient to use only one \eEDM{} measurement, the  JILA \HfF.
We take into account three representative cases that illustrate the importance of the different coupling constants; 
$\phi$ interacts \caseA~only with protons, $\gS^n=\gP^n=0$;
\caseB~only with neutrons, $\gS^p=\gP^p=0$; and 
\caseC~equally with protons and neutrons, $\gS^p=\gS^n$ and $\gP^p=\gP^n$.
The resulting 90\,\%~confidence limit~(C.L.) upper bounds for $m_\phi \ll \keV $ are given in Table~\ref{tbl:results}, and bounds as a function of $\mNP$ are plotted in Fig.~\ref{fig:onlyNN}. 
This results in the strongest terrestrial bounds, to the best of our knowledge. 
For example, the bounds from the proton and neutron EDMs are weaker by at least six orders of magnitude.
However, astrophysical bounds from stellar cooling are stronger by 3 or 4 orders of magnitude but are subject to different systematics and are model dependent, see the above discussion. 

\begin{figure}[t]
    \centering
    \includegraphics[width=\columnwidth]{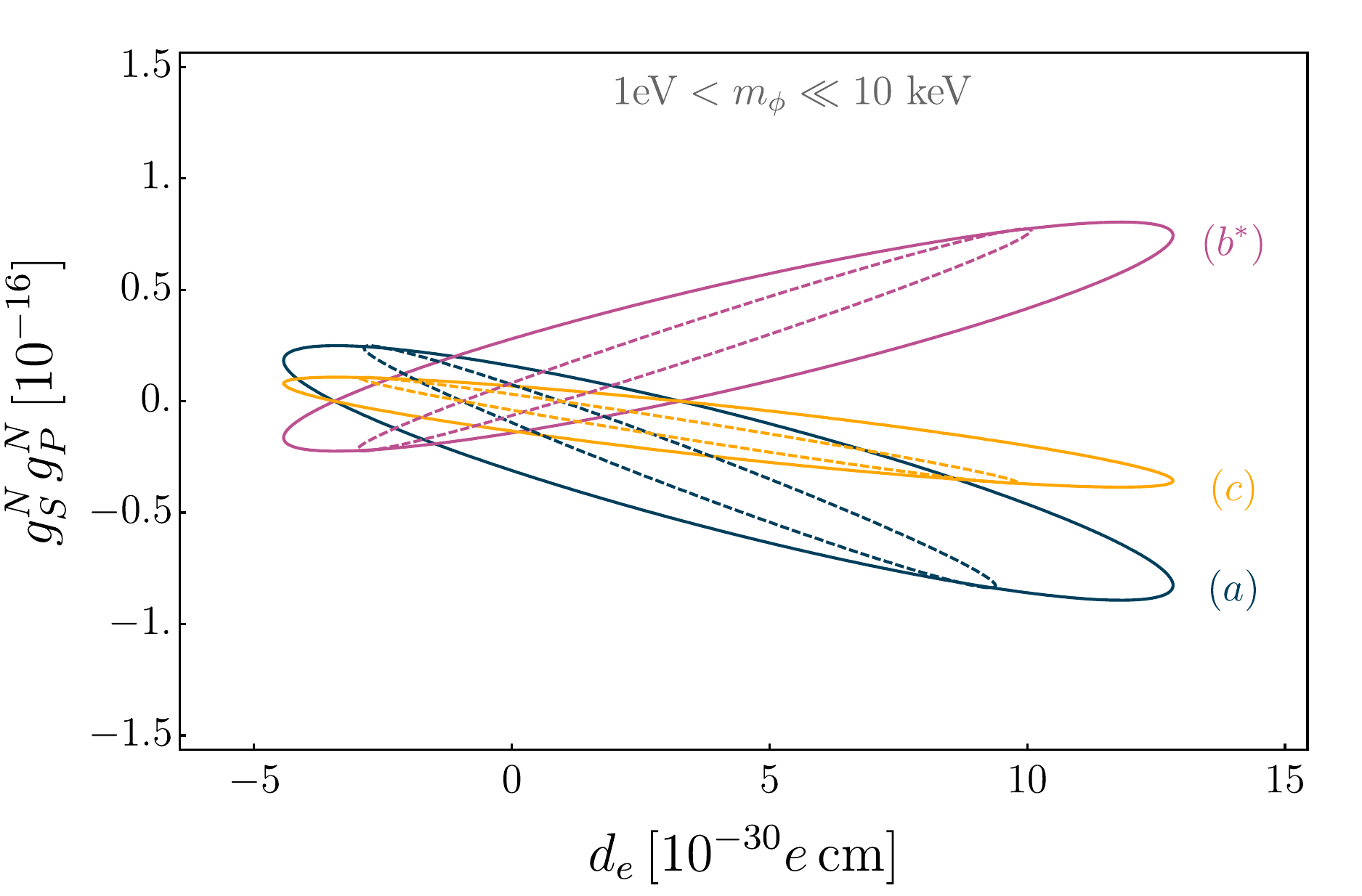}
    \caption{
    The allowed region in the $d_e$--$NN$ plane, when setting the electron-nucleon interaction to zero ($\alpha_\phi^{ie} = 0$), for $1 \eV < m_\phi \ll 10\,\eV[k]$. 
    The solid lines are derived from measurements of \HfF and ThO, and the dotted lines are derived from the measurement of \HfF and the projection of \ThF. 
    All three cases are plotted here, as indicated in the plot. 
    Note that for~\caseB, we rescaled the contour by a factor of 10 along the $\gS^N \gP^N$ axis.}
    \label{fig:deNN}
\end{figure}

Second, we consider $d_e$ and $NN$ as CPV sources and assume vanishing $\phi$--$e$ couplings. 
In this scenario, we probe $d_e$ and $\gS^N \gP^N$.
Profiling with respect to each parameter and taking the $m_\phi\ll10\,\keV$ limit, the 90\%~CL intervals are given in Table~\ref{tbl:results}.
The allowed region in the $\gS^N \gP^N$--$d_e$ plane is shown in Fig.~\ref{fig:deNN}, with solid lines for the measurement result and dashed lines for the projection. 
The translated bounds for the aforementioned cases, \caseA, \caseB, \caseC, are given in Table~\ref{tbl:results}. 

Third, we consider three CPV sources, $d_e$, $NN$ and $eN$. 
For simplicity, we take the limit of $m_\phi \ll \keV$, generalization for higher mass is straightforward. 
Profiling each time on the other parameter, the 90\%~CL intervals are given in Table~\ref{tbl:results}.
The allowed regions in the $\gS^{n,p}\gP^e$--$d_e$, $\gS^{n,p}\gP^{n,p}$--$d_e$ and $\gS^{n,p}\gP^e$--$\gS^{n,p}\gP^{n,p}$ planes can be found in the Supplemental Material.

\mysec{Conclusions}
In this work, we point out that the \eEDM{} searches in diatomic molecules, composed of at least one nucleus with a nonvanishing spin, are sensitive to new long range CPV nucleon-nucleon interactions, thus, proposing a new set of NP models that can be probed with \eEDM{} searches..  
Based on the \eEDM{} searches in \HfF, ThO, and YbF, we have derived novel bounds on such a hypothetical interaction mediated by a spin-0 particle with scalar and pseudo-scalar couplings to nucleons. 
The resulting bounds are the most stringent obtained from laboratory experiments, improving existing limits by 6 orders of magnitude in the $1\,\eV-10\,\keV$ mass range. 
For example, we show that in the presence of $eN$ and $NN$ the bound on the $d_e$ will be weaker by 2 orders of magnitude compared to the simple case of only \eEDM{} as a source of CPV. 
We project that adding the expected \ThF measurement will improve the sensitivity by about two orders of magnitude. 

There are two questions left for future work which we note.
For \eEDM{} searches using polyatomic molecules that rely on $l$-doubling, the sensitivity to the $NN$ effect needs to be determined  from the direction of the nuclear spin, \eg~\cite{Kozyryev:2017cwq,Augenbraun:2019hpx}.
\eEDM{} searches on excited rotational and vibrational states can vary the magnitude of the $NN$ interaction contribution providing more information within a single molecule to isolate the effect.

\begin{acknowledgments}
We gratefully thank Yevgeny Stadnik for careful reading of the manuscript and fruitful discussions. 
We also thank Gilad Perez and Tomer Volansky for useful discussions. 
We thank Eric Cornell, Gilad Perez, Jun Ye for comments on the manuscript. 
C.B. and Y.So. are supported by Grants from the NSF-BSF (Grant No. 2021800), the ISF (Grant No. 483/20), the BSF (Grant No. 2020300) and by the Azrieli foundation. 
Y.Sh. gratefully acknowledges funding by the European Union (ERC, 101116032 – Q-ChiMP).
Views and opinions expressed are however those of the authors only and do not necessarily reflect those of the European Union or the European Research Council Executive Agency. Neither the European Union nor the granting authority can be
held responsible for them.
\end{acknowledgments}

\bibliographystyle{utphys28mod}
\bibliography{edm_bib.bib}

\input{supplement}

\end{document}

%% file: supplement.tex
\clearpage
\onecolumngrid

\setcounter{equation}{0}
\setcounter{figure}{0}
\setcounter{table}{0}
\setcounter{section}{0}
\makeatletter
\renewcommand{\theequation}{S\arabic{equation}}
\renewcommand{\thefigure}{S\arabic{figure}}
\renewcommand{\thetable}{S\arabic{table}}
\renewcommand{\thesection}{S\arabic{section}}
\begin{center}
 \large{\bf 
 Constraining CP violating nucleon-nucleon long range interactions in diatomic eEDM searches
 }\\
 Supplemental Material
\end{center}
\begin{center}
Chaja Baruch,
P. Bryan Changala,
Yuval Shagam,
and Yotam Soreq
\end{center}

In the supplemental material we give further details on the mapping of the nucleus couplings to the nucleon couplings. Additionally we provide contour plots for the different interactions when profiling over one each time.

\section{Nucleus to nucleon spin couplings}

The pseudoscalar coupling $\alphaNP$ contains both the spin-dependent part and the mass difference part between the proton and neutron. 
We define 
\begin{align}
	B_{p\,(n)} = \frac{\overline{m}_N}{m_{p\,(n)}} b_{p\,(n)}
\end{align}
where $\overline{m}_N$ is the average nucleon mass, $m_{p\,(n)}$ is the proton\,(neutron) mass and $b_{p(n)}$ is the effective pseudoscalar proton\,(neutron) coupling of the nucleus. 
Following~\cite{JacksonKimball:2014vsz}, this effective coupling $b_{p\,(n)}$ is obtained by parameterizing the spin couplings to new physics in terms of an exotic atomic dipole moment, which is related to the coupling constants for the electron, proton, and neutron. 
The total atomic angular momentum is related to the electron spin ($S$) and nuclear spin ($I$) via the Russell-Saunders LS-coupling scheme, $\expval{F}=\expval{S} + \expval{L} + \expval{I}$, and can then relate the exotic atom dipole moment to the coupling constants 
\begin{align}
	g_A = g_e \frac{\expval{S\cdot F}}{F(F+1)} + g_N \frac{\expval{I\cdot F}}{F(F+1)}\,.
\end{align}
The coupling constant of the nucleus $g_N$ is related to the proton and neutron couplings $g_p$, and $g_n$. 
Generally, there are four contributions to the nuclear spin: proton and neutron spin and orbital angular momentum
\begin{align}
	I &= b_p I + b_n I + b_{\ell,p} + b_{\ell,n} \nonumber \\
	&= \expval{I_p} + \expval{I_n} + \expval{L_p} + \expval{L_n}
\end{align}
for $b_p + b_n + b_{\ell,p} + b_{\ell,n} = 1$. 
The exotic dipole moment coupling constant thus is $g_N = b_p g_p + b_n g_n$. 

As a first approximation (the Schmidt model) it is assumed the nucleus spin is due only to the orbital motion and spin of a single nucleon (the valence nucleon) and the spin and angular momentum of all other nucleons sums to zero
\begin{align}
	b^{(0)}_s =
	\begin{cases}
		\frac{1}{2\ell+1}, & \text{ for } I = \ell + \frac{1}{2} \\
		\frac{-1}{2\ell+1}, & \text{ for } I = \ell - \frac{1}{2}
	\end{cases}
\end{align}
and $b^{(0)}_\ell = 1-b^{(0)}_s$. 
Here, $s=n \text{ or } p$, depending on whether the atom has a valence neutron or proton. 
When measuring nuclear spin-dependent properties, the contribution of non-valence nucleon spins cannot be neglected due to the polarization of these nucleons by the valence nucleon~\cite{Stadnik:2014xja}. 
The preferred Flambaum-Tedesco~(FT) model accounts for this by using the Schmidt model as a first approximation, via $b_s^{(0)} = b_p + b_n, b^{(0)}_\ell = b_{\ell,p} + b_{\ell,n}$, and then assuming that the total angular momentum of protons is conserved separately from neutrons, such that 
\begin{align}
	b_p + b_{\ell p} &= 
	\begin{cases}
		1 &\text{ for valence proton,} \\ 
		0 &\text{ for valence neutron,}
	\end{cases} \nonumber \\ 
	b_p &= 
	\begin{cases}
		\frac{G_I - G_n b_s^{(0)}-1}{G_p - G_n - 1} &\text{ for valence proton,} \\
		\frac{G_I - G_n b_s^{(0)}}{G_p - G_n - 1} &\text{ for valence neutron,}
	\end{cases} \nonumber \\
	b_n &= b^{(0)}_s - b_p\,,
\end{align}
where $G$ are the spin g-factors.
For all molecules discussed in this work, only $^{19}$F carries spin, and its effective pseudoscalar coupling is
\begin{align}\label{eq:gPF}
	g^{\rm F}_P = 1.08 \frac{\overline{m}_N}{m_p} g_p^p - 0.08 \frac{\overline{m}_N}{m_n} g_n^p\,,
\end{align}
where we used the fact that $^{19}$F has a valence neutron, and $g_{p,n}^{p}$ are the pseudoscalar proton and neutron couplings, respectively. 
To obtain this result, we used $G_{\rm F} = 5.26$ for the spin g-factor of $^{19}$F.

\section{contour plots}

\begin{figure*}[h]
	\includegraphics[width=0.3\textwidth]{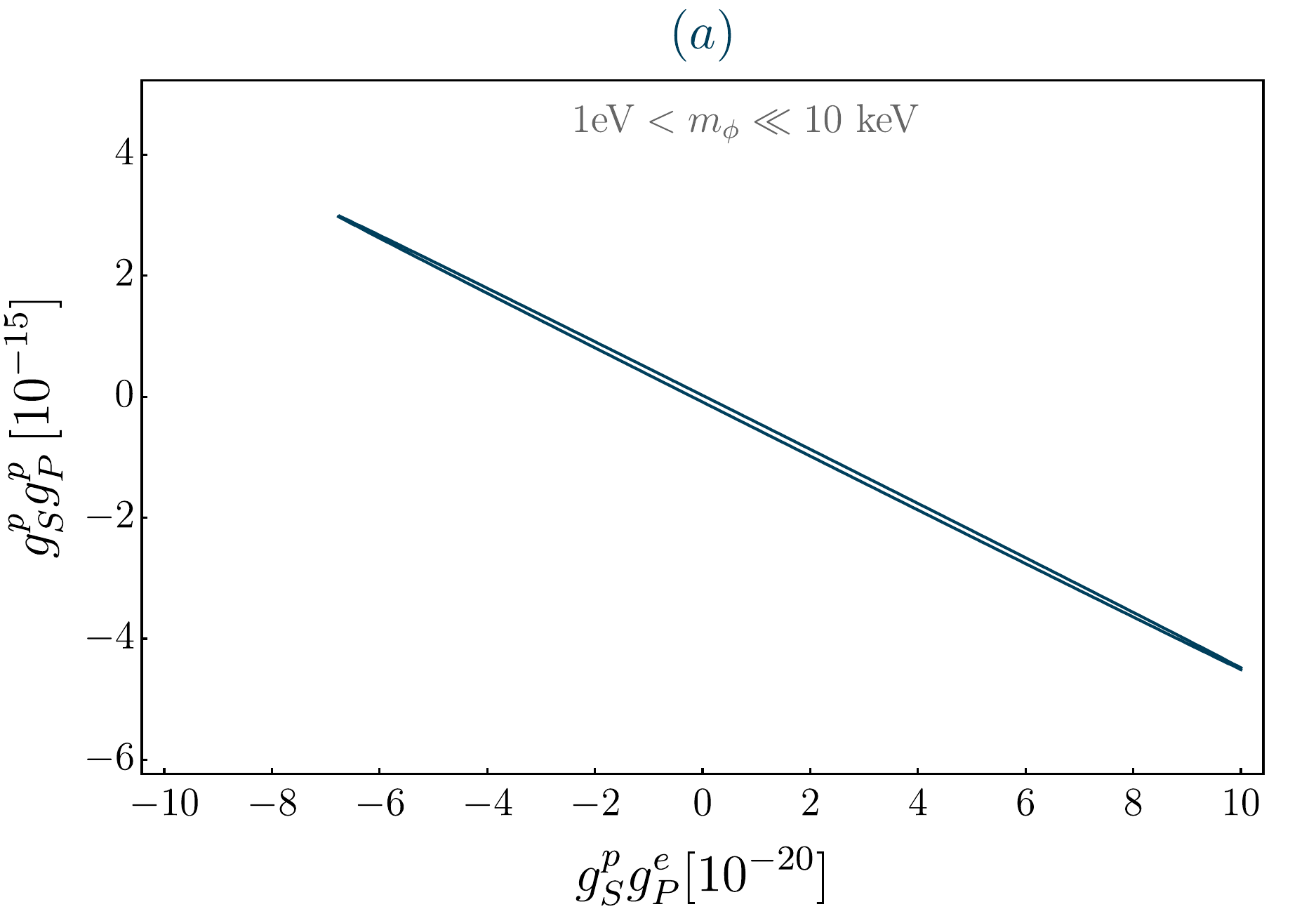}
	\includegraphics[width=0.3\textwidth]{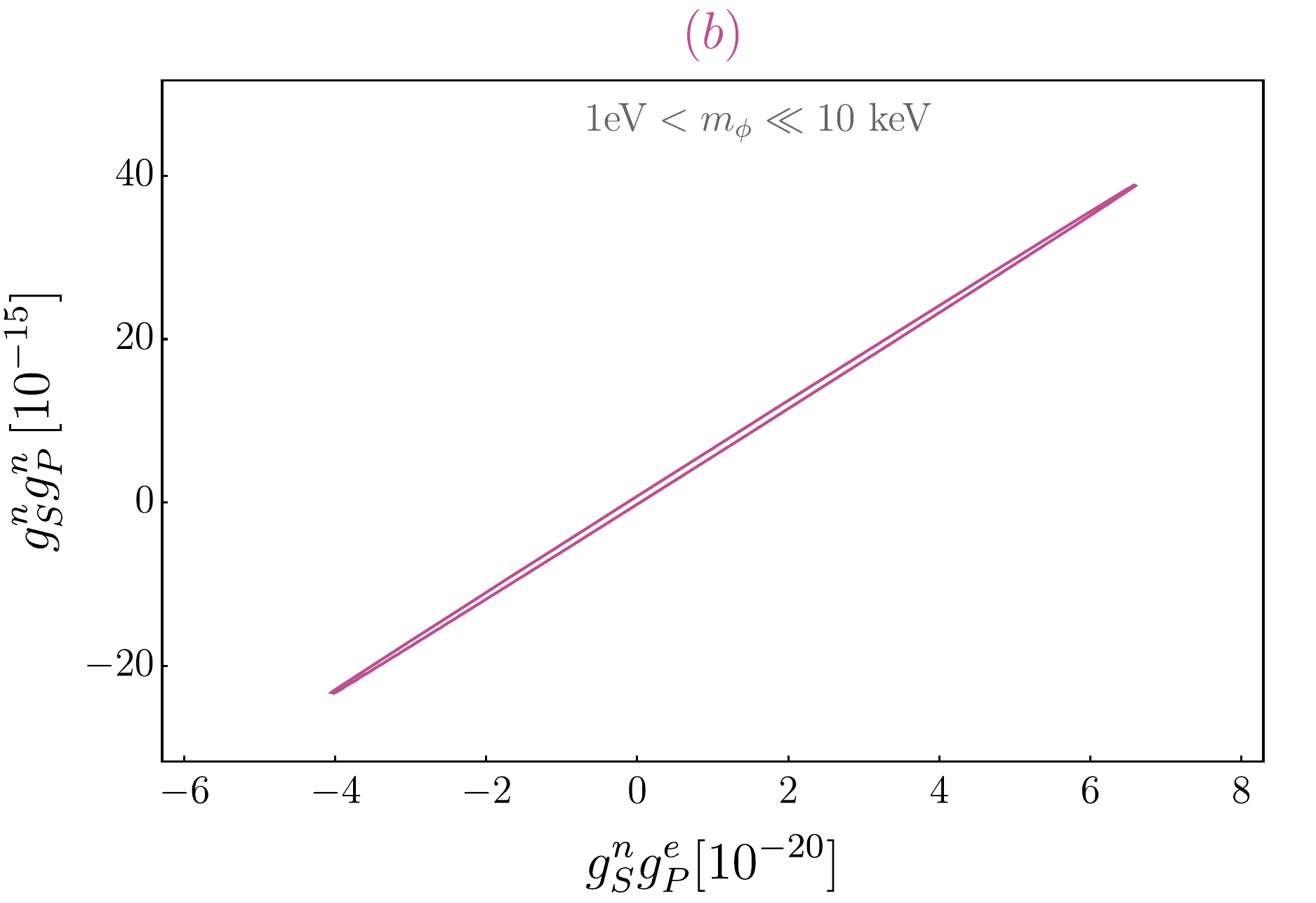}
	\includegraphics[width=0.3\textwidth]{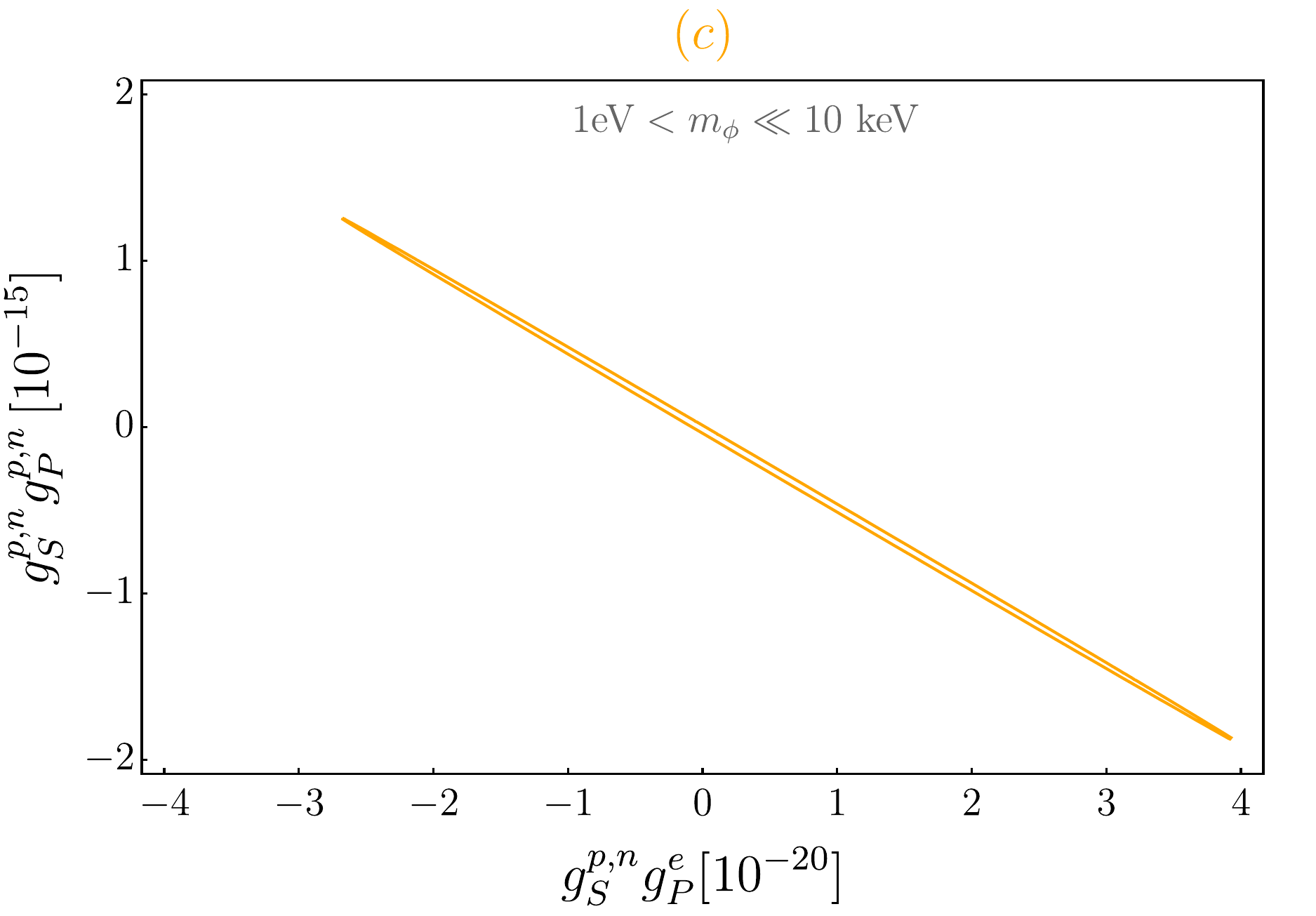}
	\includegraphics[width=0.3\textwidth]{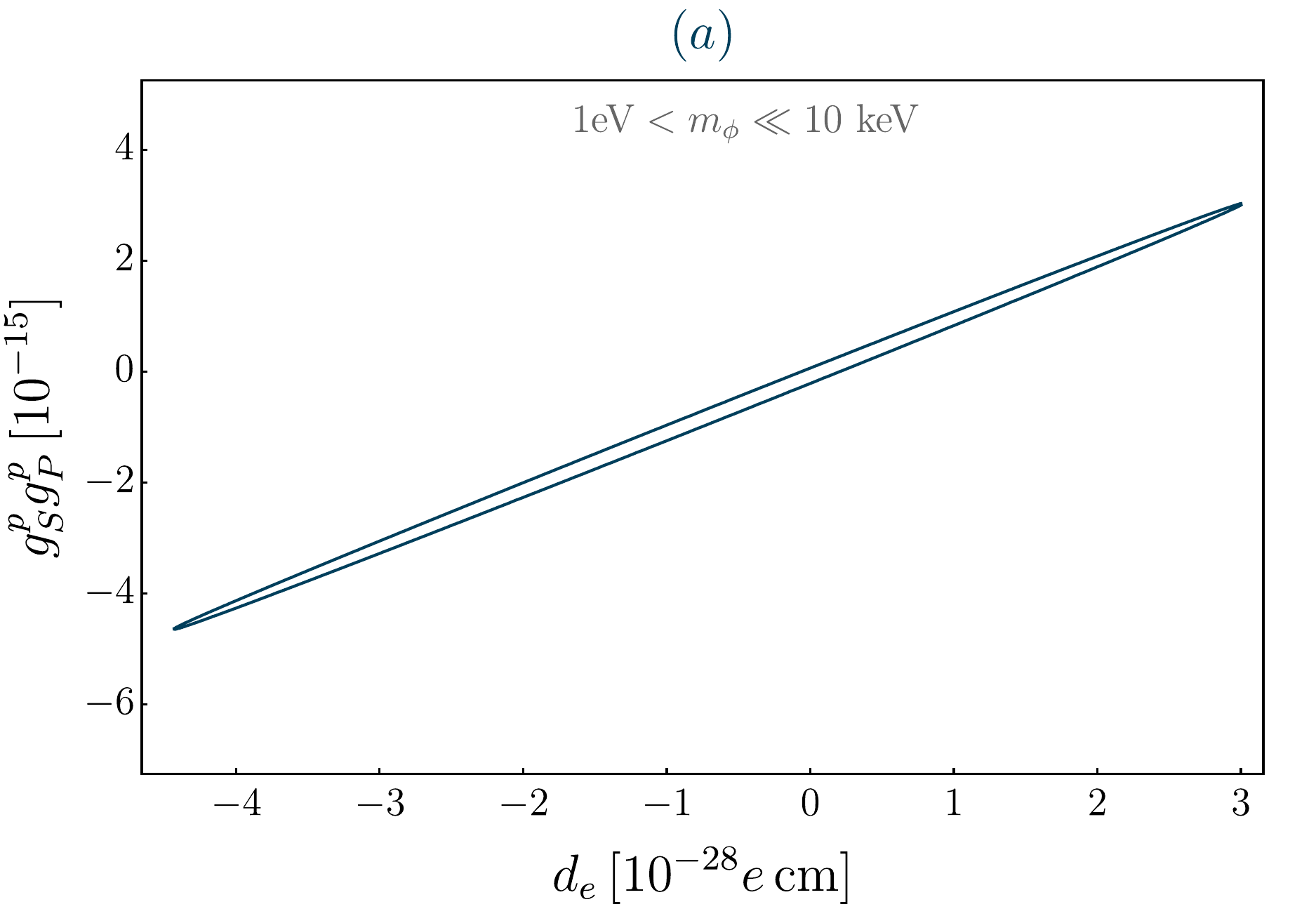}
	\includegraphics[width=0.3\textwidth]{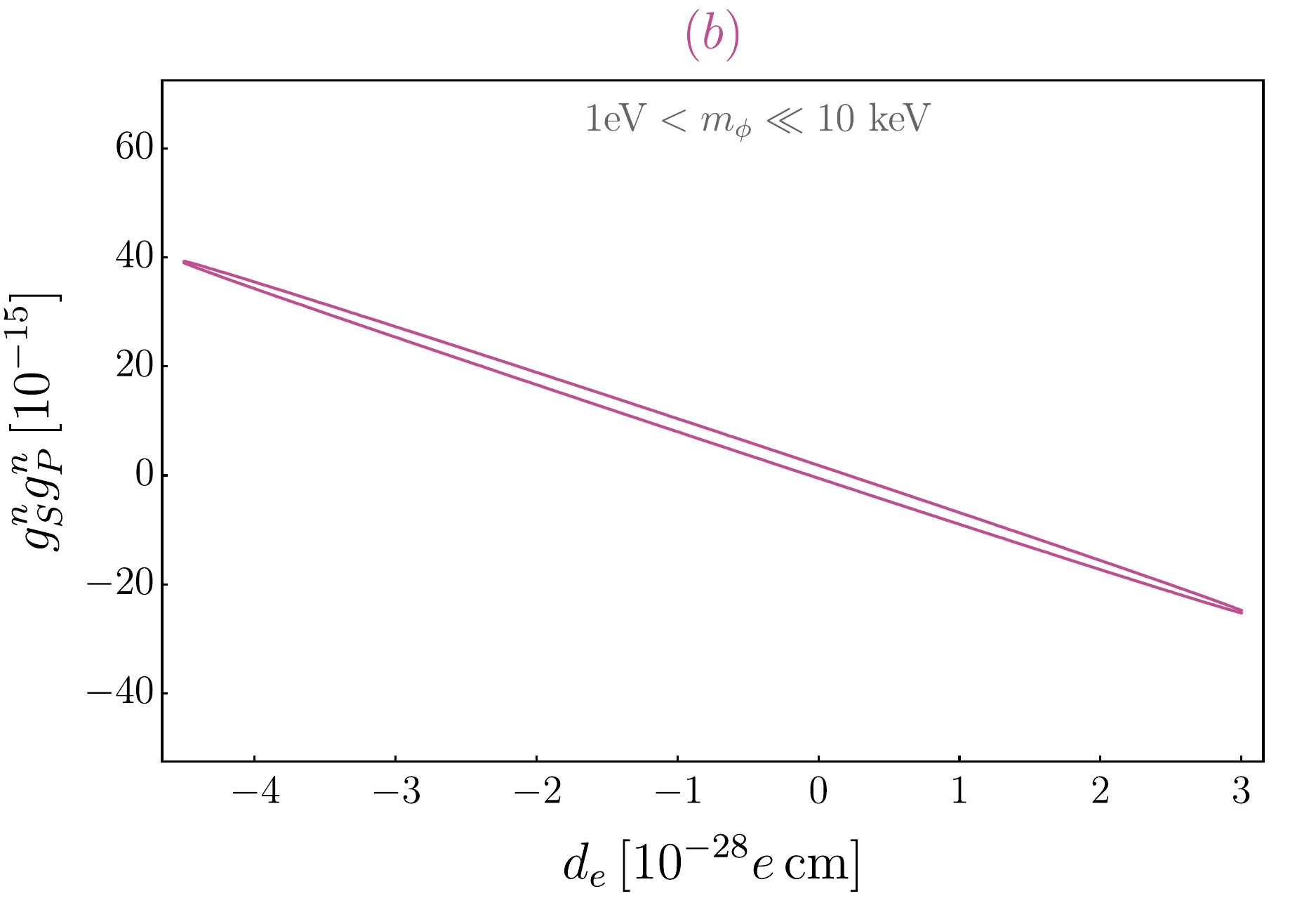}
	\includegraphics[width=0.3\textwidth]{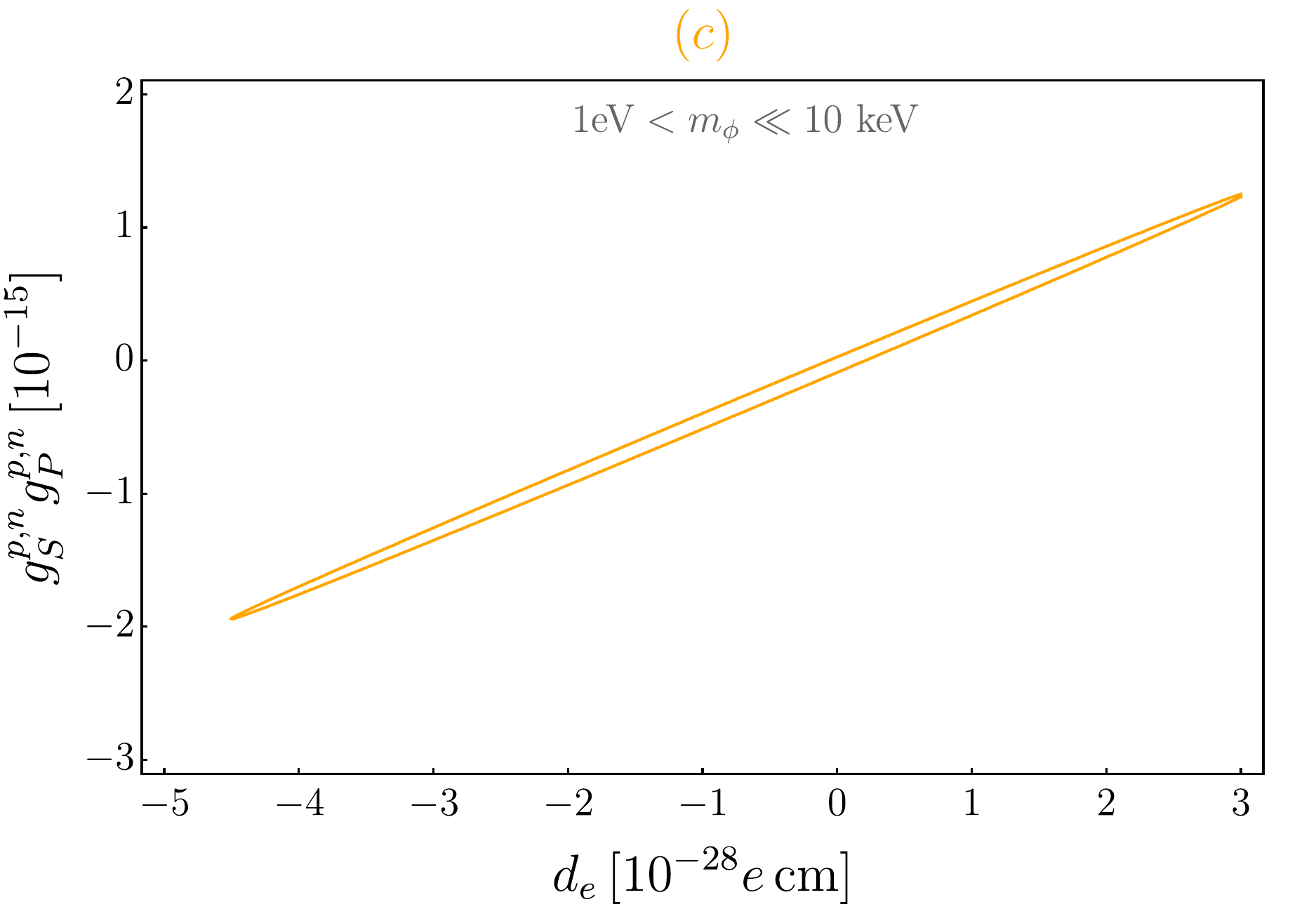}
	\includegraphics[width=0.3\textwidth]{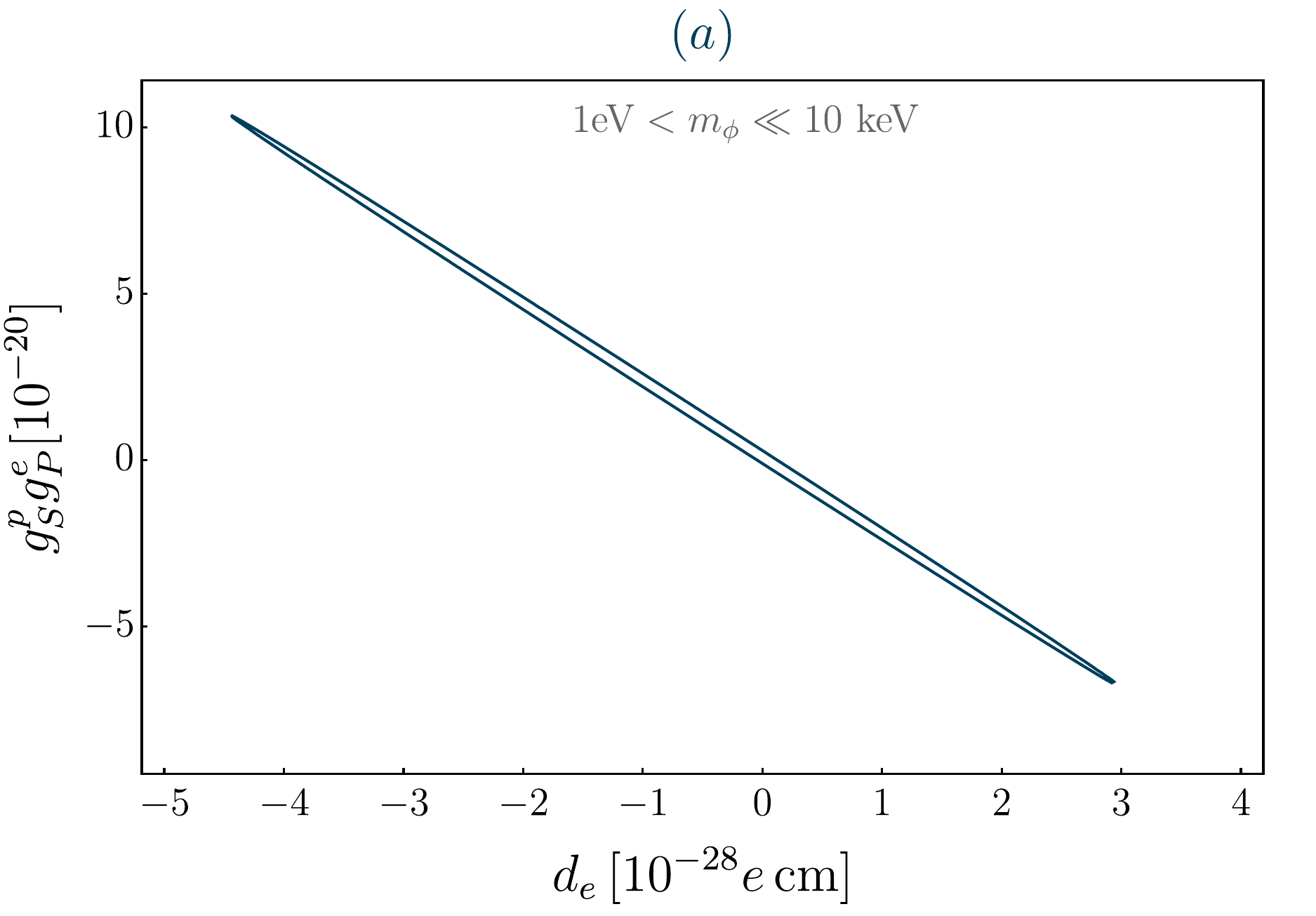}
	\includegraphics[width=0.3\textwidth]{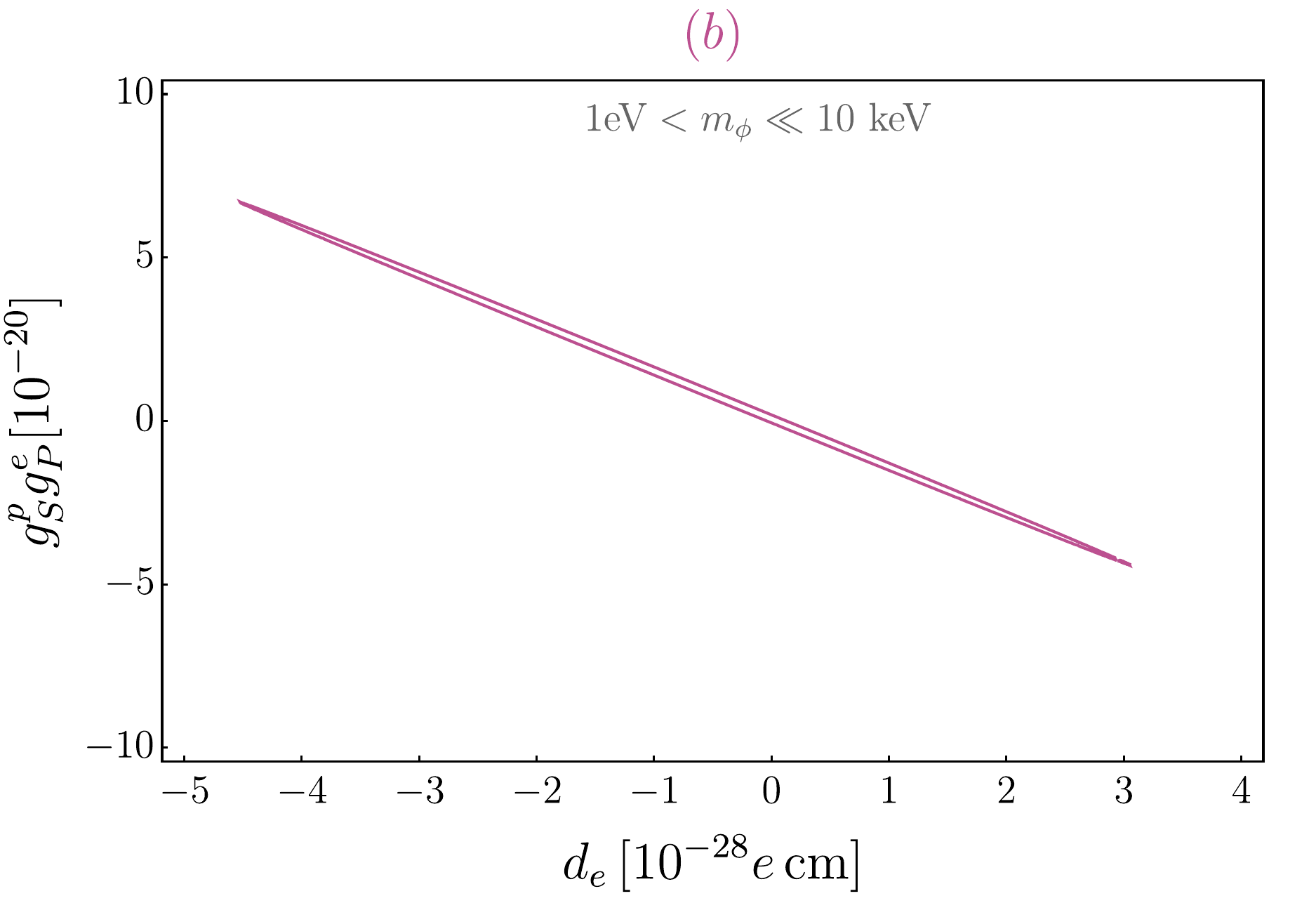}
	\includegraphics[width=0.3\textwidth]{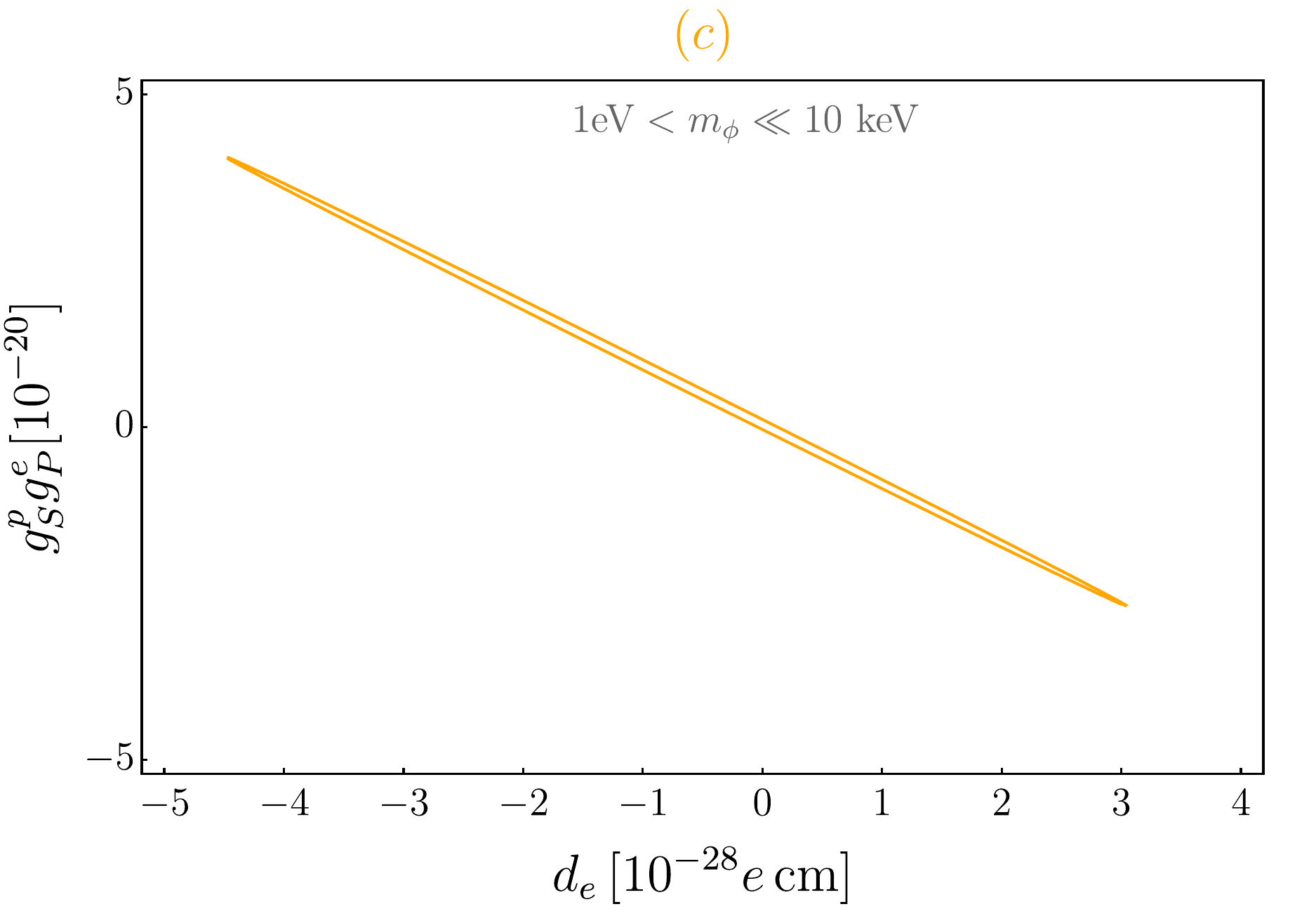}
	\caption{Contour plots for $NN$, $eN$ when profiling over $d_e$ (first row), for $NN$, $d_e$ when profiling over $Ne$ (second row), and for $d_e$, $eN$ when profiling over $NN$ (third row). All profiling is done for $1 \eV < m_\phi \ll 10 \eV[k]$.
		The results are given for the three cases \caseA, \caseB{} and \caseC{} as indicated.}
	\label{fig:deeNNN}
\end{figure*}

When considering three CPV sources, $d_e$, $NN$ and $eN$, in the limit of $m_\phi \ll \keV$, we profile each time on the other parameter, and give the 90\%~CL intervals in Table~II in the main text.
The allowed regions in the $\gS^{n,p}\gP^e$--$d_e$, $\gS^{n,p}\gP^{n,p}$--$d_e$ and $\gS^{n,p}\gP^e$--$\gS^{n,p}\gP^{n,p}$ planes are shown in Fig.~\ref{fig:deeNNN}.

%% file: NN_eEDM_PRL_v2.bbl
\providecommand{\href}[2]{#2}\begingroup\raggedright\begin{thebibliography}{10}

\bibitem{EuropeanStrategyforParticlePhysicsPreparatoryGroup:2019qin}
R.~K.~Ellis {\em et~al.}, ``{Physics Briefing Book}: {Input for the European
  Strategy for Particle Physics Update 2020}.'' {\ttfamily
  \href{https://arxiv.org/abs/1910.11775}{arXiv:1910.11775}}.

\bibitem{Safronova:2017xyt}
M.~S.~Safronova, {\em et al.}, ``{Search for New Physics with Atoms and
  Molecules},'' \href{https://dx.doi.org/10.1103/RevModPhys.90.025008}{Rev.\
  Mod.\  Phys.\  {\bfseries 90} (2018) 025008} {\ttfamily
  [\href{https://arxiv.org/abs/1710.01833}{arXiv:1710.01833}]}.

\bibitem{Pospelov:1997uv}
M.~Pospelov, ``{CP odd interaction of axion with matter},''
  \href{https://dx.doi.org/10.1103/PhysRevD.58.097703}{Phys.\  Rev.\  D
  {\bfseries 58} (1998) 097703} {\ttfamily
  [\href{https://arxiv.org/abs/hep-ph/9707431}{hep-ph/9707431}]}.

\bibitem{Ginges:2003qt}
J.~S.~M.~Ginges and V.~V.~Flambaum, ``{Violations of fundamental symmetries in
  atoms and tests of unification theories of elementary particles},''
  \href{https://dx.doi.org/10.1016/j.physrep.2004.03.005}{Phys.\  Rept.\
  {\bfseries 397} (2004) 63--154} {\ttfamily
  [\href{https://arxiv.org/abs/physics/0309054}{physics/0309054}]}.

\bibitem{Pospelov:2005pr}
M.~Pospelov and A.~Ritz, ``{Electric dipole moments as probes of new
  physics},'' \href{https://dx.doi.org/10.1016/j.aop.2005.04.002}{Annals Phys.\
   {\bfseries 318} (2005) 119--169} {\ttfamily
  [\href{https://arxiv.org/abs/hep-ph/0504231}{hep-ph/0504231}]}.

\bibitem{Engel:2013lsa}
J.~Engel, M.~J.~Ramsey-Musolf, and U.~van Kolck, ``{Electric Dipole Moments of
  Nucleons, Nuclei, and Atoms: The Standard Model and Beyond},''
  \href{https://dx.doi.org/10.1016/j.ppnp.2013.03.003}{Prog.\  Part.\  Nucl.\
  Phys.\  {\bfseries 71} (2013) 21--74} {\ttfamily
  [\href{https://arxiv.org/abs/1303.2371}{arXiv:1303.2371}]}.

\bibitem{Chupp:2017rkp}
T.~E.~Chupp, P.~Fierlinger, M.~J.~Ramsey-Musolf, and J.~T.~Singh, ``{Electric
  dipole moments of atoms, molecules, nuclei, and particles},''
  \href{https://dx.doi.org/10.1103/RevModPhys.91.015001}{Rev.\  Mod.\  Phys.\
  {\bfseries 91} (2019) 015001} {\ttfamily
  [\href{https://arxiv.org/abs/1710.02504}{arXiv:1710.02504}]}.

\bibitem{Gaul:2023hdd}
K.~Gaul and R.~Berger, ``{Global analysis of CP-violation in atoms, molecules
  and role of medium-heavy systems}.'' {\ttfamily
  \href{https://arxiv.org/abs/2312.08858}{arXiv:2312.08858}}.

\bibitem{Stadnik:2017hpa}
Y.~V.~Stadnik, V.~A.~Dzuba, and V.~V.~Flambaum, ``{Improved Limits on
  Axionlike-Particle-Mediated P , T -Violating Interactions between Electrons
  and Nucleons from Electric Dipole Moments of Atoms and Molecules},''
  \href{https://dx.doi.org/10.1103/PhysRevLett.120.013202}{Phys.\  Rev.\
  Lett.\  {\bfseries 120} (2018) 013202} {\ttfamily
  [\href{https://arxiv.org/abs/1708.00486}{arXiv:1708.00486}]}.

\bibitem{Fleig:2018bsf}
T.~Fleig and M.~Jung, ``{Model-independent determinations of the electron EDM
  and the role of diamagnetic atoms},''
  \href{https://dx.doi.org/10.1007/JHEP07(2018)012}{JHEP {\bfseries 07} (2018)
  012} {\ttfamily [\href{https://arxiv.org/abs/1802.02171}{arXiv:1802.02171}]}.

\bibitem{Prosnyak:2023duq}
S.~D.~Prosnyak, D.~E.~Maison, and L.~V.~Skripnikov, ``{Updated Constraints on
  ,-Violating Axionlike-Particle-Mediated Electron\textendash{}Electron and
  Electron\textendash{}Nucleus Interactions from HfF$^{+}$ Experiment},''
  \href{https://dx.doi.org/10.3390/sym15051043}{Symmetry {\bfseries 15} (2023)
  1043} {\ttfamily
  [\href{https://arxiv.org/abs/2304.07164}{arXiv:2304.07164}]}.

\bibitem{Flambaum:2019ejc}
V.~V.~Flambaum, M.~Pospelov, A.~Ritz, and Y.~V.~Stadnik, ``{Sensitivity of EDM
  experiments in paramagnetic atoms and molecules to hadronic CP violation},''
  \href{https://dx.doi.org/10.1103/PhysRevD.102.035001}{Phys.\  Rev.\  D
  {\bfseries 102} (2020) 035001} {\ttfamily
  [\href{https://arxiv.org/abs/1912.13129}{arXiv:1912.13129}]}.

\bibitem{Roussy:2022cmp}
T.~S.~Roussy {\em et~al.}, ``{An improved bound on the
  electron\textquoteright{}s electric dipole moment},''
  \href{https://dx.doi.org/10.1126/science.adg4084}{Science {\bfseries 381}
  (2023) adg4084} {\ttfamily
  [\href{https://arxiv.org/abs/2212.11841}{arXiv:2212.11841}]}.

\bibitem{Caldwell:2022xwj}
L.~Caldwell {\em et~al.}, ``{Systematic and statistical uncertainty evaluation
  of the HfF+ electron electric dipole moment experiment},''
  \href{https://dx.doi.org/10.1103/PhysRevA.108.012804}{Phys.\  Rev.\  A
  {\bfseries 108} (2023) 012804} {\ttfamily
  [\href{https://arxiv.org/abs/2212.11837}{arXiv:2212.11837}]}.

\bibitem{ACME:2018yjb}
{\bfseries ACME} Collaboration, ``{Improved limit on the electric dipole moment
  of the electron},''
  \href{https://dx.doi.org/10.1038/s41586-018-0599-8}{Nature {\bfseries 562}
  (2018) 355--360}.

\bibitem{Hudson:2011zz}
J.~J.~Hudson, {\em et al.}, ``{Improved measurement of the shape of the
  electron},'' \href{https://dx.doi.org/10.1038/nature10104}{Nature {\bfseries
  473} (2011) 493--496}.

\bibitem{Kara:2012ay}
D.~M.~Kara, {\em et al.}, ``{Measurement of the electron's electric dipole
  moment using YbF molecules: methods and data analysis},''
  \href{https://dx.doi.org/10.1088/1367-2630/14/10/103051}{New J.\  Phys.\
  {\bfseries 14} (2012) 103051} {\ttfamily
  [\href{https://arxiv.org/abs/1208.4507}{arXiv:1208.4507}]}.

\bibitem{Ng:2022fzm}
K.~B.~Ng {\em et~al.}, ``{Spectroscopy on the
  electron-electric-dipole-moment\textendash{}sensitive states of ThF+},''
  \href{https://dx.doi.org/10.1103/PhysRevA.105.022823}{Phys.\  Rev.\  A
  {\bfseries 105} (2022) 022823} {\ttfamily
  [\href{https://arxiv.org/abs/2202.01346}{arXiv:2202.01346}]}.

\bibitem{Bottaro:2023gep}
S.~Bottaro, A.~Caputo, G.~Raffelt, and E.~Vitagliano, ``{Stellar limits on
  scalars from electron-nucleus bremsstrahlung},''
  \href{https://dx.doi.org/10.1088/1475-7516/2023/07/071}{JCAP {\bfseries 07}
  (2023) 071} {\ttfamily
  [\href{https://arxiv.org/abs/2303.00778}{arXiv:2303.00778}]}.

\bibitem{Hardy:2016kme}
E.~Hardy and R.~Lasenby, ``{Stellar cooling bounds on new light particles:
  plasma mixing effects},''
  \href{https://dx.doi.org/10.1007/JHEP02(2017)033}{JHEP {\bfseries 02} (2017)
  033} {\ttfamily [\href{https://arxiv.org/abs/1611.05852}{arXiv:1611.05852}]}.

\bibitem{Buschmann:2021juv}
M.~Buschmann, C.~Dessert, J.~W.~Foster, A.~J.~Long, and B.~R.~Safdi, ``{Upper
  Limit on the QCD Axion Mass from Isolated Neutron Star Cooling},''
  \href{https://dx.doi.org/10.1103/PhysRevLett.128.091102}{Phys.\  Rev.\
  Lett.\  {\bfseries 128} (2022) 091102} {\ttfamily
  [\href{https://arxiv.org/abs/2111.09892}{arXiv:2111.09892}]}.

\bibitem{OHare:2020wah}
C.~A.~J.~O'Hare and E.~Vitagliano, ``{Cornering the axion with $CP$-violating
  interactions},'' \href{https://dx.doi.org/10.1103/PhysRevD.102.115026}{Phys.\
   Rev.\  D {\bfseries 102} (2020) 115026} {\ttfamily
  [\href{https://arxiv.org/abs/2010.03889}{arXiv:2010.03889}]}.

\bibitem{Moody:1984ba}
J.~E.~Moody and F.~Wilczek, ``{NEW MACROSCOPIC FORCES?}''
  \href{https://dx.doi.org/10.1103/PhysRevD.30.130}{Phys.\  Rev.\  D {\bfseries
  30} (1984) 130}.

\bibitem{DiLuzio:2023lmd}
L.~Di~Luzio, H.~Gisbert, G.~Levati, P.~Paradisi, and P.~S\o{}rensen,
  ``{CP-Violating Axions: A Theory Review}.'' {\ttfamily
  \href{https://arxiv.org/abs/2312.17310}{arXiv:2312.17310}}.

\bibitem{DiLuzio:2023edk}
L.~Di~Luzio, G.~Levati, and P.~Paradisi, ``{The Chiral Lagrangian of
  CP-Violating Axion-Like Particles}.'' {\ttfamily
  \href{https://arxiv.org/abs/2311.12158}{arXiv:2311.12158}}.

\bibitem{Okawa:2021fto}
S.~Okawa, M.~Pospelov, and A.~Ritz, ``{Long-range axion forces and hadronic CP
  violation},'' \href{https://dx.doi.org/10.1103/PhysRevD.105.075003}{Phys.\
  Rev.\  D {\bfseries 105} (2022) 075003} {\ttfamily
  [\href{https://arxiv.org/abs/2111.08040}{arXiv:2111.08040}]}.

\bibitem{Dekens:2022gha}
W.~Dekens, J.~de~Vries, and S.~Shain, ``{CP-violating axion interactions in
  effective field theory},''
  \href{https://dx.doi.org/10.1007/JHEP07(2022)014}{JHEP {\bfseries 07} (2022)
  014} {\ttfamily [\href{https://arxiv.org/abs/2203.11230}{arXiv:2203.11230}]}.

\bibitem{Graham:2015cka}
P.~W.~Graham, D.~E.~Kaplan, and S.~Rajendran, ``{Cosmological Relaxation of the
  Electroweak Scale},''
  \href{https://dx.doi.org/10.1103/PhysRevLett.115.221801}{Phys.\  Rev.\
  Lett.\  {\bfseries 115} (2015) 221801} {\ttfamily
  [\href{https://arxiv.org/abs/1504.07551}{arXiv:1504.07551}]}.

\bibitem{Gupta:2015uea}
R.~S.~Gupta, Z.~Komargodski, G.~Perez, and L.~Ubaldi, ``{Is the Relaxion an
  Axion?}'' \href{https://dx.doi.org/10.1007/JHEP02(2016)166}{JHEP {\bfseries
  02} (2016) 166} {\ttfamily
  [\href{https://arxiv.org/abs/1509.00047}{arXiv:1509.00047}]}.

\bibitem{Flacke:2016szy}
T.~Flacke, C.~Frugiuele, E.~Fuchs, R.~S.~Gupta, and G.~Perez, ``{Phenomenology
  of relaxion-Higgs mixing},''
  \href{https://dx.doi.org/10.1007/JHEP06(2017)050}{JHEP {\bfseries 06} (2017)
  050} {\ttfamily [\href{https://arxiv.org/abs/1610.02025}{arXiv:1610.02025}]}.

\bibitem{AxionLimits}
C.~O'Hare, ``cajohare/AxionLimits: AxionLimits.''
  \url{https://cajohare.github.io/AxionLimits/}, 2020.
\newblock \href{https://dx.doi.org/10.5281/zenodo.3932430}{{\ttfamily
  doi:10.5281/zenodo.3932430}}.

\bibitem{Carenza:2019pxu}
P.~Carenza, {\em et al.}, ``{Improved axion emissivity from a supernova via
  nucleon-nucleon bremsstrahlung},''
  \href{https://dx.doi.org/10.1088/1475-7516/2019/10/016}{JCAP {\bfseries 10}
  (2019) 016} {\ttfamily
  [\href{https://arxiv.org/abs/1906.11844}{arXiv:1906.11844}]}. [Erratum: JCAP
  05, E01 (2020)].

\bibitem{Bar:2019ifz}
N.~Bar, K.~Blum, and G.~D'Amico, ``{Is there a supernova bound on axions?}''
  \href{https://dx.doi.org/10.1103/PhysRevD.101.123025}{Phys.\  Rev.\  D
  {\bfseries 101} (2020) 123025} {\ttfamily
  [\href{https://arxiv.org/abs/1907.05020}{arXiv:1907.05020}]}.

\bibitem{Burrage:2016bwy}
C.~Burrage and J.~Sakstein, ``{A Compendium of Chameleon Constraints},''
  \href{https://dx.doi.org/10.1088/1475-7516/2016/11/045}{JCAP {\bfseries 11}
  (2016) 045} {\ttfamily
  [\href{https://arxiv.org/abs/1609.01192}{arXiv:1609.01192}]}.

\bibitem{Masso:2005ym}
E.~Masso and J.~Redondo, ``{Evading astrophysical constraints on axion-like
  particles},'' \href{https://dx.doi.org/10.1088/1475-7516/2005/09/015}{JCAP
  {\bfseries 09} (2005) 015} {\ttfamily
  [\href{https://arxiv.org/abs/hep-ph/0504202}{hep-ph/0504202}]}.

\bibitem{Jaeckel:2006xm}
J.~Jaeckel, E.~Masso, J.~Redondo, A.~Ringwald, and F.~Takahashi, ``{The Need
  for purely laboratory-based axion-like particle searches},''
  \href{https://dx.doi.org/10.1103/PhysRevD.75.013004}{Phys.\  Rev.\  D
  {\bfseries 75} (2007) 013004} {\ttfamily
  [\href{https://arxiv.org/abs/hep-ph/0610203}{hep-ph/0610203}]}.

\bibitem{DeRocco:2020xdt}
W.~DeRocco, P.~W.~Graham, and S.~Rajendran, ``{Exploring the robustness of
  stellar cooling constraints on light particles},''
  \href{https://dx.doi.org/10.1103/PhysRevD.102.075015}{Phys.\  Rev.\  D
  {\bfseries 102} (2020) 075015} {\ttfamily
  [\href{https://arxiv.org/abs/2006.15112}{arXiv:2006.15112}]}.

\bibitem{Budnik:2020nwz}
R.~Budnik, H.~Kim, O.~Matsedonskyi, G.~Perez, and Y.~Soreq, ``{Probing the
  relaxed relaxion and Higgs portal scenarios with XENON1T scintillation and
  ionization data},''
  \href{https://dx.doi.org/10.1103/PhysRevD.104.015012}{Phys.\  Rev.\  D
  {\bfseries 104} (2021) 015012} {\ttfamily
  [\href{https://arxiv.org/abs/2006.14568}{arXiv:2006.14568}]}.

\bibitem{Bloch:2020uzh}
I.~M.~Bloch, {\em et al.}, ``{Exploring new physics with O(keV) electron
  recoils in direct detection experiments},''
  \href{https://dx.doi.org/10.1007/JHEP01(2021)178}{JHEP {\bfseries 01} (2021)
  178} {\ttfamily [\href{https://arxiv.org/abs/2006.14521}{arXiv:2006.14521}]}.

\bibitem{Escribano:2020wua}
P.~Escribano and A.~Vicente, ``{Ultralight scalars in leptonic observables},''
  \href{https://dx.doi.org/10.1007/JHEP03(2021)240}{JHEP {\bfseries 03} (2021)
  240} {\ttfamily [\href{https://arxiv.org/abs/2008.01099}{arXiv:2008.01099}]}.

\bibitem{Dolan:2022kul}
M.~J.~Dolan, F.~J.~Hiskens, and R.~R.~Volkas, ``{Advancing globular cluster
  constraints on the axion-photon coupling},''
  \href{https://dx.doi.org/10.1088/1475-7516/2022/10/096}{JCAP {\bfseries 10}
  (2022) 096} {\ttfamily
  [\href{https://arxiv.org/abs/2207.03102}{arXiv:2207.03102}]}.

\bibitem{Sahoo:2016zvr}
B.~K.~Sahoo, ``{Improved limits on the hadronic and semihadronic $CP$ violating
  parameters and role of a dark force carrier in the electric dipole moment of
  $^{199}$Hg},'' \href{https://dx.doi.org/10.1103/PhysRevD.95.013002}{Phys.\
  Rev.\  D {\bfseries 95} (2017) 013002} {\ttfamily
  [\href{https://arxiv.org/abs/1612.09371}{arXiv:1612.09371}]}.

\bibitem{Abel:2020pzs}
C.~Abel {\em et~al.}, ``{Measurement of the Permanent Electric Dipole Moment of
  the Neutron},''
  \href{https://dx.doi.org/10.1103/PhysRevLett.124.081803}{Phys.\  Rev.\
  Lett.\  {\bfseries 124} (2020) 081803} {\ttfamily
  [\href{https://arxiv.org/abs/2001.11966}{arXiv:2001.11966}]}.

\bibitem{Pendlebury:2015lrz}
J.~M.~Pendlebury {\em et~al.}, ``{Revised experimental upper limit on the
  electric dipole moment of the neutron},''
  \href{https://dx.doi.org/10.1103/PhysRevD.92.092003}{Phys.\  Rev.\  D
  {\bfseries 92} (2015) 092003} {\ttfamily
  [\href{https://arxiv.org/abs/1509.04411}{arXiv:1509.04411}]}.

\bibitem{Mantry:2014zsa}
S.~Mantry, M.~Pitschmann, and M.~J.~Ramsey-Musolf, ``{Distinguishing axions
  from generic light scalars using electric dipole moment and fifth-force
  experiments},'' \href{https://dx.doi.org/10.1103/PhysRevD.90.054016}{Phys.\
  Rev.\  D {\bfseries 90} (2014) 054016} {\ttfamily
  [\href{https://arxiv.org/abs/1401.7339}{arXiv:1401.7339}]}.

\bibitem{Lee:2018vaq}
J.~Lee, A.~Almasi, and M.~Romalis, ``{Improved Limits on Spin-Mass
  Interactions},''
  \href{https://dx.doi.org/10.1103/PhysRevLett.120.161801}{Phys.\  Rev.\
  Lett.\  {\bfseries 120} (2018) 161801} {\ttfamily
  [\href{https://arxiv.org/abs/1801.02757}{arXiv:1801.02757}]}.

\bibitem{Jenke:2012ju}
T.~Jenke, {\em et al.}, ``{A quantized frequency reference in the short-ranged
  gravity potential and its application for dark matter and dark energy
  searches}.'' {\ttfamily
  \href{https://arxiv.org/abs/1208.3875}{arXiv:1208.3875}}.

\bibitem{Tullney:2013wqa}
K.~Tullney {\em et~al.}, ``{Constraints on Spin-Dependent Short-Range
  Interaction between Nucleons},''
  \href{https://dx.doi.org/10.1103/PhysRevLett.111.100801}{Phys.\  Rev.\
  Lett.\  {\bfseries 111} (2013) 100801} {\ttfamily
  [\href{https://arxiv.org/abs/1303.6612}{arXiv:1303.6612}]}.

\bibitem{Guigue:2015fyt}
M.~Guigue, D.~Jullien, A.~K.~Petukhov, and G.~Pignol, ``{Constraining
  short-range spin-dependent forces with polarized $^3$He},''
  \href{https://dx.doi.org/10.1103/PhysRevD.92.114001}{Phys.\  Rev.\  D
  {\bfseries 92} (2015) 114001} {\ttfamily
  [\href{https://arxiv.org/abs/1511.06993}{arXiv:1511.06993}]}.

\bibitem{Feng:2022tsu}
Y.~K.~Feng, D.~H.~Ning, S.~B.~Zhang, Z.~T.~Lu, and D.~Sheng, ``{Search for
  Monopole-Dipole Interactions at the Submillimeter Range with a Xe129-Xe131-Rb
  Comagnetometer},''
  \href{https://dx.doi.org/10.1103/PhysRevLett.128.231803}{Phys.\  Rev.\
  Lett.\  {\bfseries 128} (2022) 231803} {\ttfamily
  [\href{https://arxiv.org/abs/2205.13237}{arXiv:2205.13237}]}.

\bibitem{SchillerHD1}
S.~Alighanbari, G.~S.~Giri, F.~L.~Constantin, V.~I.~Korobov, and S.~Schiller,
  ``Precise test of quantum electrodynamics and determination of fundamental
  constants with HD+ ions,''
  \href{https://dx.doi.org/10.1038/s41586-020-2261-5}{Nature {\bfseries 581}
  (2020) 152--158}.

\bibitem{SchillerHD2}
S.~Alighanbari, I.~V.~Kortunov, G.~S.~Giri, and S.~Schiller, ``Test of charged
  baryon interaction with high-resolution vibrational spectroscopy of molecular
  hydrogen ions,'' \href{https://dx.doi.org/10.1038/s41567-023-02088-2}{Nature
  Physics {\bfseries 19} (2023) 1263--1269}.

\bibitem{Goudzovski:2022vbt}
E.~Goudzovski {\em et~al.}, ``{New physics searches at kaon and hyperon
  factories},'' \href{https://dx.doi.org/10.1088/1361-6633/ac9cee}{Rept.\
  Prog.\  Phys.\  {\bfseries 86} (2023) 016201} {\ttfamily
  [\href{https://arxiv.org/abs/2201.07805}{arXiv:2201.07805}]}.

\bibitem{Nesvizhevsky:2007by}
V.~V.~Nesvizhevsky, G.~Pignol, and K.~V.~Protasov, ``{Neutron scattering and
  extra short range interactions},''
  \href{https://dx.doi.org/10.1103/PhysRevD.77.034020}{Phys.\  Rev.\  D
  {\bfseries 77} (2008) 034020} {\ttfamily
  [\href{https://arxiv.org/abs/0711.2298}{arXiv:0711.2298}]}.

\bibitem{Kamiya:2015eva}
Y.~Kamiya, K.~Itagaki, M.~Tani, G.~N.~Kim, and S.~Komamiya, ``{Constraints on
  New Gravitylike Forces in the Nanometer Range},''
  \href{https://dx.doi.org/10.1103/PhysRevLett.114.161101}{Phys.\  Rev.\
  Lett.\  {\bfseries 114} (2015) 161101} {\ttfamily
  [\href{https://arxiv.org/abs/1504.02181}{arXiv:1504.02181}]}.

\bibitem{Haddock:2017wav}
C.~C.~Haddock {\em et~al.}, ``{Search for deviations from the inverse square
  law of gravity at nm range using a pulsed neutron beam},''
  \href{https://dx.doi.org/10.1103/PhysRevD.97.062002}{Phys.\  Rev.\  D
  {\bfseries 97} (2018) 062002} {\ttfamily
  [\href{https://arxiv.org/abs/1712.02984}{arXiv:1712.02984}]}.

\bibitem{Heacock:2021btd}
B.~Heacock {\em et~al.}, ``{Pendell\"osung interferometry probes the neutron
  charge radius, lattice dynamics, and fifth forces},''
  \href{https://dx.doi.org/10.1126/science.abc2794}{Science {\bfseries 373}
  (2021) abc2794} {\ttfamily
  [\href{https://arxiv.org/abs/2103.05428}{arXiv:2103.05428}]}.

\bibitem{Ledbetter:2012xd}
M.~P.~Ledbetter, M.~V.~Romalis, and D.~F.~Jackson-Kimball, ``{Constraints on
  short-range spin-dependent interactions from scalar spin-spin coupling in
  deuterated molecular hydrogen},''
  \href{https://dx.doi.org/10.1103/PhysRevLett.110.040402}{Phys.\  Rev.\
  Lett.\  {\bfseries 110} (2013) 040402} {\ttfamily
  [\href{https://arxiv.org/abs/1203.6894}{arXiv:1203.6894}]}.

\bibitem{Mostepanenko:2020lqe}
V.~M.~Mostepanenko, A.~A.~Starobinsky, and E.~N.~Velichko, eds., ``{The State
  of the Art in Constraining Axion-to-Nucleon Coupling and Non-Newtonian
  Gravity from Laboratory Experiments},''
  \href{https://dx.doi.org/10.3390/universe6090147}{Universe {\bfseries 6}
  (2020) 147} {\ttfamily
  [\href{https://arxiv.org/abs/2009.04517}{arXiv:2009.04517}]}.

\bibitem{Gorghetto:2021luj}
M.~Gorghetto, G.~Perez, I.~Savoray, and Y.~Soreq, ``{Probing CP violation in
  photon self-interactions with cavities},''
  \href{https://dx.doi.org/10.1007/JHEP10(2021)056}{JHEP {\bfseries 10} (2021)
  056} {\ttfamily [\href{https://arxiv.org/abs/2103.06298}{arXiv:2103.06298}]}.

\bibitem{Dobrescu:2006au}
B.~A.~Dobrescu and I.~Mocioiu, ``{Spin-dependent macroscopic forces from new
  particle exchange},''
  \href{https://dx.doi.org/10.1088/1126-6708/2006/11/005}{JHEP {\bfseries 11}
  (2006) 005} {\ttfamily
  [\href{https://arxiv.org/abs/hep-ph/0605342}{hep-ph/0605342}]}.

\bibitem{JacksonKimball:2014vsz}
D.~F.~Jackson~Kimball, ``{Nuclear spin content and constraints on exotic
  spin-dependent couplings},''
  \href{https://dx.doi.org/10.1088/1367-2630/17/7/073008}{New J.\  Phys.\
  {\bfseries 17} (2015) 073008} {\ttfamily
  [\href{https://arxiv.org/abs/1407.2671}{arXiv:1407.2671}]}.

\bibitem{Fadeev:2018rfl}
P.~Fadeev, {\em et al.}, ``{Revisiting spin-dependent forces mediated by new
  bosons: Potentials in the coordinate-space representation for macroscopic-
  and atomic-scale experiments},''
  \href{https://dx.doi.org/10.1103/PhysRevA.99.022113}{Phys.\  Rev.\  A
  {\bfseries 99} (2019) 022113} {\ttfamily
  [\href{https://arxiv.org/abs/1810.10364}{arXiv:1810.10364}]}.

\bibitem{Flambaum:2006ip}
V.~V.~Flambaum and A.~F.~Tedesco, ``{Dependence of nuclear magnetic moments on
  quark masses and limits on temporal variation of fundamental constants from
  atomic clock experiments},''
  \href{https://dx.doi.org/10.1103/PhysRevC.73.055501}{Phys.\  Rev.\  C
  {\bfseries 73} (2006) 055501} {\ttfamily
  [\href{https://arxiv.org/abs/nucl-th/0601050}{nucl-th/0601050}]}.

\bibitem{Stadnik:2014xja}
Y.~V.~Stadnik and V.~V.~Flambaum, ``{Nuclear spin-dependent interactions:
  Searches for WIMP, Axion and Topological Defect Dark Matter, and Tests of
  Fundamental Symmetries},''
  \href{https://dx.doi.org/10.1140/epjc/s10052-015-3326-8}{Eur.\  Phys.\  J.\
  C {\bfseries 75} (2015) 110} {\ttfamily
  [\href{https://arxiv.org/abs/1408.2184}{arXiv:1408.2184}]}.

\bibitem{PhysRevA.76.030501}
A.~N.~Petrov, N.~S.~Mosyagin, T.~A.~Isaev, and A.~V.~Titov, ``Theoretical study
  of $\mathrm{Hf}{\mathrm{F}}^{+}$ in search of the electron electric dipole
  moment,'' \href{https://dx.doi.org/10.1103/PhysRevA.76.030501}{Phys.\  Rev.\
  A {\bfseries 76} (2007) 030501(R)}.

\bibitem{Khristenko1998}
S.~V.~Khristenko, V.~P.~Shevelko, and A.~I.~Maslov, {\em Molecular Structure},
  \href{https://dx.doi.org/10.1007/978-3-642-71946-2_1}{pp.~1--19}.
\newblock Springer Berlin Heidelberg, Berlin, Heidelberg, 1998.

\bibitem{Barker2012}
B.~J.~Barker, I.~O.~Antonov, M.~C.~Heaven, and K.~A.~Peterson, ``{Spectroscopic
  investigations of ThF and ThF$^+$},''
  \href{https://dx.doi.org/10.1063/1.3691301}{The Journal of Chemical Physics
  {\bfseries 136} (2012) 104305}.

\bibitem{PhysRevA.91.042504}
L.~V.~Skripnikov and A.~V.~Titov, ``Theoretical study of ${\mathrm{ThF}}^{+}$
  in the search for $T,P$-violation effects: Effective state of a Th atom in
  ${\mathrm{ThF}}^{+}$ and ThO compounds,''
  \href{https://dx.doi.org/10.1103/PhysRevA.91.042504}{Phys.\  Rev.\  A
  {\bfseries 91} (2015) 042504}.

\bibitem{Dzuba:2018anu}
V.~A.~Dzuba, V.~V.~Flambaum, I.~B.~Samsonov, and Y.~V.~Stadnik, ``{New
  constraints on axion-mediated P,T-violating interaction from electric dipole
  moments of diamagnetic atoms},''
  \href{https://dx.doi.org/10.1103/PhysRevD.98.035048}{Phys.\  Rev.\  D
  {\bfseries 98} (2018) 035048} {\ttfamily
  [\href{https://arxiv.org/abs/1805.01234}{arXiv:1805.01234}]}.

\bibitem{kbthesis}
K.~B.~Ng, {\em The ThF$^+$ eEDM experiment: concept, design, and
  characterization}.
\newblock Phd thesis, University of Colorado, Boulder, CO, 2024.

\bibitem{Kozyryev:2017cwq}
I.~Kozyryev and N.~R.~Hutzler, ``{Precision Measurement of Time-Reversal
  Symmetry Violation with Laser-Cooled Polyatomic Molecules},''
  \href{https://dx.doi.org/10.1103/PhysRevLett.119.133002}{Phys.\  Rev.\
  Lett.\  {\bfseries 119} (2017) 133002} {\ttfamily
  [\href{https://arxiv.org/abs/1705.11020}{arXiv:1705.11020}]}.

\bibitem{Augenbraun:2019hpx}
B.~L.~Augenbraun, {\em et al.}, ``{Laser-Cooled Polyatomic Molecules for
  Improved Electron Electric Dipole Moment Searches},''
  \href{https://dx.doi.org/10.1088/1367-2630/ab687b}{New J.\  Phys.\
  {\bfseries 22} (2020) 022003} {\ttfamily
  [\href{https://arxiv.org/abs/1910.11318}{arXiv:1910.11318}]}.

\end{thebibliography}\endgroup
